\documentclass[11pt,a4paper]{article}
\usepackage{etex}
\usepackage[thinlines]{easybmat}
\usepackage[T1]{fontenc}
\usepackage{pstricks-add}
\usepackage[body={16cm,23cm}]{geometry}
\usepackage{enumerate}
\usepackage{amsmath,amssymb}
\usepackage{bm}
\usepackage{epic}
\usepackage{graphicx}
\usepackage{color}
\usepackage{hyperref}

\newcommand{\SpinAux}{{\bf J}}

\newcommand{\Identity}{\mathbb{I}}

\newcommand{\specbaz}{z}

\newcommand{\Qop}{{\rm \bf{Q}}}

\newcommand{\Top}{{\rm {\bf T}}}

\newcommand{\Lop}{{\mathcal L}}

\newcommand{\Lbb}{{\mathbb L}}
\newcommand{\Lbf}{{\mathbf L}}

\newcommand{\Dbb}{{\mathbb D}}
\newcommand{\Dbf}{{\mathcal D}}
\newcommand{\Rbf}{{\mathbf R}}
\newcommand{\Pbf}{{\mathbf P}}
\newcommand{\Xbf}{{\mathbf X}}
\newcommand{\beq}{\begin{equation}}
\newcommand{\eeq}{\end{equation}}

\newcommand{\Qf}{ {\bf Q}}

\newcommand{\alg}[1]{\mathfrak{#1}}

\newcommand{\oscalg}{{\mathcal H}}

\newcommand{\oscb}{\mathbf{b}}

\newcommand{\osch}{\mathbf{h}}

\newcommand{\sfrac}[2]{{\textstyle\frac{#1}{#2}}}
\newcommand{\half}{\sfrac{1}{2}}

\newcommand{\Gbb}{{\mathbb G}}
\newcommand{\Fbb}{{\mathbb F}}

\newcommand{\E}{{\mathcal J}}
\newcommand{\oE}{\overline{\mathcal J}}
\newcommand{\oJ}{\overline{\mathbf J}}
\newcommand{\oscbp}{{\mathbf b}^{\dagger}}
\newcommand{\oscbm}{{\mathbf b}^{\phantom{\dagger}}}%
\newcommand{\osccp}{{\mathbf c}^{\dagger}}
\newcommand{\osccm}{{\mathbf c}^{\phantom{\dagger}}}%

\newcommand{\oscbpo}{{\mathbf b}^{{\dagger}\,[1]}}
\newcommand{\oscbmo}{{\mathbf b}^{[1]\phantom{\dagger}} }
\newcommand{\oscbpt}{{\mathbf b}^{{\dagger}\,[2]}}
\newcommand{\oscbmt}{{\mathbf b}^{[2]\phantom{\dagger}} }
\newcommand{\p}{{p}}

\renewcommand{\a}{{a}}
\renewcommand{\b}{{b}}
\newcommand{\g}{{c}}

\newcommand{\e}{e}

\newcommand{\ac}{{\textsc a}}
\newcommand{\bc}{{\textsc b}}
\newcommand{\gc}{{\textsc c}}

\newcommand{\rep}{{\Lambda}}
\newcommand{\m}{{\lambda}}

\newcommand{\ds}{{\displaystyle}}
\newcommand{\bea}{\begin{eqnarray}}
\newcommand{\eea}{\end{eqnarray}}

\newcommand{\ntr}{{\rm \widehat{Tr}}}

\makeatletter
\def\mr@ignsp#1 {\ifx\:#1\@empty\else #1\expandafter\mr@ignsp\fi}%
\newcommand{\multiref}[1]{\begingroup
\xdef\mr@no@sparg{\expandafter\mr@ignsp#1 \: }%
\def\mr@comma{}%
\@for\mr@refs:=\mr@no@sparg\do{\mr@comma\def\mr@comma{,}\ref{\mr@refs}}%
\endgroup}
\makeatother

\numberwithin{equation}{section}

\begin{document}
\psset{arrowscale=2}
\thispagestyle{empty}
\pagenumbering{alph}

\begin{flushright}\footnotesize
\texttt{HU-Mathematik:~2010-9}\\
\texttt{HU-EP-10/28}\\
\texttt{AEI-2010-116}\\
\vspace{0.5cm}
\end{flushright}
\setcounter{footnote}{0}

\begin{center}
{\Large\textbf{\mathversion{bold}
Baxter Q-Operators and Representations of Yangians
}\par}
\vspace{15mm}

{\sc Vladimir V.~Bazhanov $^{a,b}$, Rouven Frassek $^{c,e}$, Tomasz {\L}ukowski 
$^{c,d}$,\\ Carlo Meneghelli $^{c,e}$,
Matthias Staudacher $^{c,e}$}\\[5mm]

{\it $^a$ Department of Theoretical Physics,
    Research School of Physics and Engineering\\
    Australian National University, Canberra, ACT 0200, Australia}\\[5mm]

{\it $^b$ Mathematical Sciences Institute,\\
      Australian National University, Canberra, ACT 0200, Australia}\\[5mm]

{\it $^c$ Institut f\"ur Mathematik und Institut f\"ur Physik,
 Humboldt-Universit\"at zu Berlin\\  
Johann von Neumann-Haus, Rudower Chaussee 25, 12489 Berlin, Germany
}\\[5mm]

{\it $^d$ Institute of Physics, Jagellonian University\\
    ul.~Reymonta 4, 30-059 Krak\'ow, Poland}\\[5mm]

{\it $^e$ Max-Planck-Institut f\"ur Gravitationsphysik,
 Albert-Einstein-Institut\\ 
   Am M\"uhlenberg 1, 14476 Potsdam, Germany}\\[5mm]

\texttt{Vladimir.Bazhanov@anu.edu.au}\\
\texttt{rfrassek@physik.hu-berlin.de}\\
\texttt{lukowski@mathematik.hu-berlin.de}\\
\texttt{carlo@aei.mpg.de}\\
\texttt{matthias@aei.mpg.de}\\[18mm]

\textbf{Abstract}\\[2mm]
\end{center}

\noindent{We develop a new approach to Baxter ${\bf Q}$-operators by relating
them to the theory of Yangians, which are the simplest examples for
quantum groups.
Here we open up a new chapter in this theory and study
certain degenerate solutions of the Yang-Baxter equation connected
with harmonic oscillator algebras.
These infinite-state solutions of the Yang-Baxter equation
serve as elementary, ``partonic'' building blocks for
other solutions via the standard fusion procedure.
As a first example of the method we consider $\alg{sl}(n)$ compact spin chains
and derive the full hierarchy of operatorial functional equations for
all related commuting transfer matrices and ${\bf Q}$-operators.
This leads to a systematic and transparent solution of these chains, where
the nested Bethe equations are derived in an entirely algebraic
fashion, without any reference to the traditional Bethe ansatz
techniques.}

\newpage
\clearpage\pagenumbering{arabic}
\setcounter{page}{2}
\section{Introduction and Overview}
\label{sec:intro}

The method of functional relations and commuting transfer matrices,
introduced by Baxter in his seminal paper \cite{Baxter:1972hz} on the
exact solution of the eight-vertex model, plays a fundamental role in
the theory of integrable quantum systems. It is based on an
explicit algebraic construction of transfer matrices,
which gives {\em a priori\/} knowledge about the
analytic properties of their eigenvalues. A central part within this
method involvess the so-called ${\bf Q}$-operators. These operators
are distinguished by the fact that zeroes of their eigenvalues precisely coincide with
the roots of a certain system of algebraic equations, which arises as a part of
the coordinate \cite{Baxter:book} or algebraic \cite{Faddeev:1996iy}
Bethe Ansatz.  

The underlying algebraic structure behind the construction of the
commuting transfer matrices, termed ${\bf T}$-matrices or ${\bf
 T}$-operators, is by now 
well understood. It is connected with the simplest representations of
quantum groups, which are closely related to the standard
finite-dimensional representations of classical Lie algebras.  
In contrast, the algebraic construction of the ${\bf Q}$-operators
appears to be a more complicated and, at the same time, more interesting
problem.   
Much progress in this direction has already been achieved from
a case-by-case study of various models, see
e.g.~\cite{Baxter:1972hz,Baxter:book,Bazhanov:1989nc,Gaudin:1992ci,
Bazhanov:1996dr,Smirnov:2000jpa,Faddeev:2000if, Derkachov:2003qb,Bytsko:2006ut,
Bazhanov:2010ts} but the problem still continues to reveal its new
features.  

In this paper we develop a new approach to ${\bf Q}$-operators by
connecting them to the theory of Yangians, which are the simplest
examples of quantum groups. In doing so we shall develop new aspects
of the theory of infinite-dimensional representations of Yangians,
naturally leading to a systematic and transparent construction of the
${\bf Q}$-operators. Here we illustrate our approach on the compact
$\alg{gl}(n)$-spin chains, but the results may be readily
generalized to other models, and in particular to supersymmetric spin
chains \cite{Frassek:2010ga}.

Let us then consider the integrable $\alg{gl}(n)$-spin chain with the
well-known Hamiltonian  
\begin{equation}
\label{sln-ham}
\mathbf{H}_n
= 
2\sum_{l=1}^L\left(1- \sum_{a,b=1}^n e_{ab}^{(l)}
\,e_{ba}^{(l+1)} \right)
\end{equation}
in the presence of twisted periodic boundary conditions,
\begin{equation}
\label{bcH}
e_{ab}^{(L+1)}:= 
e^{i\,(\Phi_a-\Phi_b)}\,e_{ab}^{(1)}
\, ,
\end{equation}
where $\Phi_1,\Phi_2,\ldots,\Phi_n$ is a set of fixed twist parameters
(or fields). Here $e_{ab}$ denotes the $n\times n$ matrix unit
$(e_{ab})_{ij}=\delta_{ai}\delta_{bj}$ and the superscript ``$(l)$''
refers to the quantum space of the $l$-th spin in the chain.  Each
``spin'' can take $n$ different values $a=1,2,\ldots,n$. It is easy to
check that for the quasi-periodic boundary conditions \eqref{bcH} the
numbers $m_1,m_2,\ldots,m_n$, counting the total number of spins of
type ``$1$'', ``$2$'', \dots, ``$n$'' in the chain, are conserved
quantum numbers for the Hamiltonian \eqref{sln-ham}. 
Due to these conservation properties this
integrable model can be solved via the ``nested'' Bethe Ansatz
technique \cite{Sutherland:1975,Lai:74}, which leads to the well known
result for the eigenvalues of \eqref{sln-ham}, see
\eqref{energy.formula} below. They are expressed through solutions of
the already mentioned algebraic equations, commonly called Bethe
Ansatz equations.

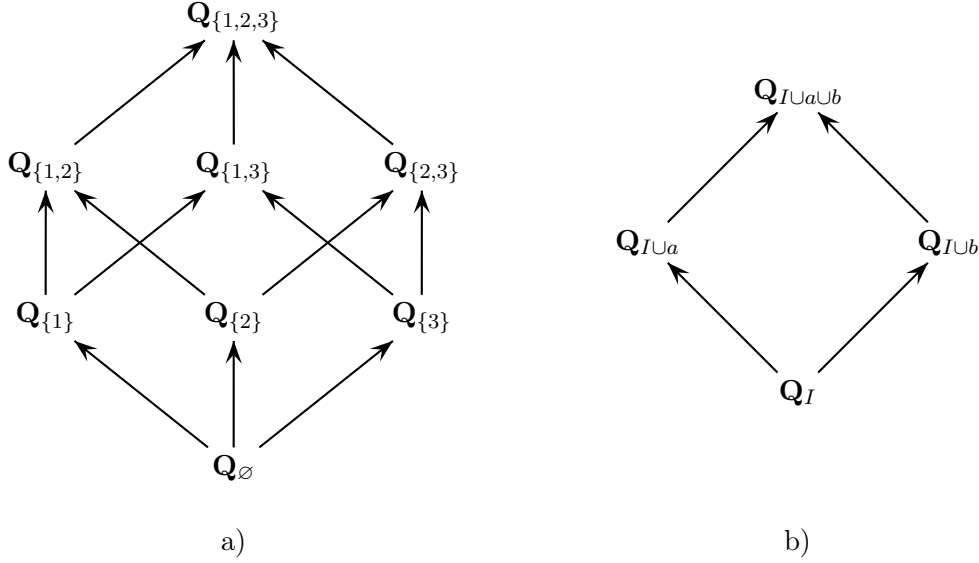
\begin{figure}[ht]
\begin{center}
\begin{pspicture}(15,7)
\rput(4,1){\rnode{A0}{$\Qf_{\varnothing}$}}
\rput(1.5,3){\rnode{A1}{$\Qf_{\{1\}}$}}
\rput(4,3){\rnode{A2}{$\Qf_{\{2\}}$}}
\rput(6.5,3){\rnode{A3}{$\Qf_{\{3\}}$}}
\rput(1.5,5){\rnode{A12}{$\Qf_{\{1,2\}}$}}
\rput(4,5){\rnode{A13}{$\Qf_{\{1,3\}}$}}
\rput(6.5,5){\rnode{A23}{$\Qf_{\{2,3\}}$}}
\rput(4,7){\rnode{A123}{$\Qf_{\{ 1,2,3\}}$}}
\ncline[ArrowInside=->,ArrowInsidePos=1.0,nodesep=0.1]{A0}{A1}
\ncline[ArrowInside=->,ArrowInsidePos=1.0,nodesep=0.1]{A0}{A2}
\ncline[ArrowInside=->,ArrowInsidePos=1.0,nodesep=0.1]{A0}{A3}
\ncline[ArrowInside=->,ArrowInsidePos=1.0,nodesep=0.1]{A1}{A12}
\ncline[ArrowInside=->,ArrowInsidePos=1.0,nodesep=0.1]{A1}{A13}
\ncline[ArrowInside=->,ArrowInsidePos=1.0,nodesep=0.1]{A2}{A12}
\ncline[ArrowInside=->,ArrowInsidePos=1.0,nodesep=0.1]{A2}{A23}
\ncline[ArrowInside=->,ArrowInsidePos=1.0,nodesep=0.1]{A3}{A13}
\ncline[ArrowInside=->,ArrowInsidePos=1.0,nodesep=0.1]{A3}{A23}
\ncline[ArrowInside=->,ArrowInsidePos=1.0,nodesep=0.1]{A12}{A123}
\ncline[ArrowInside=->,ArrowInsidePos=1.0,nodesep=0.1]{A13}{A123}
\ncline[ArrowInside=->,ArrowInsidePos=1.0,nodesep=0.1]{A23}{A123}

\rput(11.5,2){\rnode{B0}{$\Qf_I$}}
\rput(9.5,4){\rnode{B1}{$\Qf_{I\cup a}$}}
\rput(13.5,4){\rnode{B2}{$\Qf_{I\cup b}$}}
\rput(11.5,6){\rnode{B12}{$\Qf_{I\cup a\cup b}$}}
\ncline[ArrowInside=->,ArrowInsidePos=1.0,nodesep=0.1]{B0}{B1}
\ncline[ArrowInside=->,ArrowInsidePos=1.0,nodesep=0.1]{B0}{B2}
\ncline[ArrowInside=->,ArrowInsidePos=1.0,nodesep=0.1]{B1}{B12}
\ncline[ArrowInside=->,ArrowInsidePos=1.0,nodesep=0.1]{B2}{B12}
\rput(4,0){\rnode{c}{a)}}
\rput(11.5,0){\rnode{d}{b)}}
\end{pspicture}
\end{center}
\caption{a) Hasse diagram for $\mathfrak{gl}(3)$, b) An elementary
 quadrilateral, whose bottom node corresponds to the set 
$I=\{a_1,a_2,\ldots,\a_p\}$.}
\label{Hasse3}
\end{figure}

It is relatively well known that there are different but 
equivalent forms of the Bethe Ansatz. In fact, it is easy to argue
that there are precisely $n!$\  different Bethe {\em Ans\"atze} in our
case, related by all possible permutations of the occupation numbers
$m_1,m_2,\ldots,m_n$. Indeed, there are $n$ ways to choose the bare
vacuum state, then $n-1$ ways to proceed on the second ``nested''
stage of the Bethe Ansatz and so on\footnote{%
One should keep in mind, however, that the above argument only applies
to the case where all fields $\Phi_a$ take generic, non-zero
values. If some or all of the fields vanish, or else take certain
special values, only a few of the Bethe Ans\"atze are well defined,
while other ones typically suffer from multiple roots or the so-called
``beyond the equator'' problem \cite{Pronko:1998xa}.}.
These options can be conveniently depicted
by directed paths on a {\em Hasse} diagram which spans an
$n$-dimensional hypercube. The nodes of the hypercube are labeled by
increasing integer sets
$I=\{a_1,a_2,\ldots,\a_p\}\subseteq\{1,2,\ldots,n\}$,  
where $0\le p\le n$.  See the $n=3$ example in
Fig.~\ref{Hasse3}-a. There are exactly $2^n$ nodes on the diagram and
exactly $n!$ ordered paths from the bottom to the top.  Then each path
is related to a particular variant of the Bethe Ansatz, while 
nodes on that path are related to the so-called ${\rm Q}$-functions 
entering the corresponding Bethe Ansatz equations. This concise
description was proposed in \cite{Tsuboi:2009ud}. Note that the usual nested Bethe ansatz proceeds from the top to the bottom of this diagram.
\vspace{.7cm}
%

Our approach precisely reproduces this picture. We explicitly
construct the $2^n$ different ${\bf Q}$-operators, corresponding to
the nodes of the Hasse diagram, and subsequently derive the nested Bethe Ansatz
equations and eigenvalues of the Hamiltonian \eqref{sln-ham} without
any reference to the eigenvector construction.  
The different ${\bf Q}$-operators are not functionally independent. They
satisfy the so-called Hirota equations,
defined on the direct product of the
$n$-dimensional hypercube and the real line. The equations have the same form
for every quadrilateral of the Hasse diagram (see
Fig.~\ref{Hasse3}-b), 
\beq\label{quadri}
\Delta_{\{a,b\}}\Qop_{I\cup a \cup b}(\specbaz)\,\Qop_{I }(\specbaz)=
\Qop_{I \cup a}(\specbaz-\half)\,\Qop_{I \cup b}(\specbaz+\half)
-
\Qop_{I \cup b}(\specbaz-\half)\,\Qop_{I \cup a}(\specbaz+\half),
\eeq
where 
\beq\label{Delta}
\Delta_{\{a,b\}}=2\,i\,\sin\Big(\sfrac{\Phi_a-\Phi_b}{2}\Big).
\eeq
and the expression $I\cup a$ denote the union of the sets $I$ and the
one-element set $\{a\}$. The reader might be aware that the 
Hirota equations frequently arise in the analysis of  quantum
integrable models, see e.g.~\cite{Krichever:1996qd}. The
equations \eqref{quadri} for the ${\bf Q}$-operators and their
eigenvalues were gradually developed for various models related to
$\alg{gl}(n)$ algebra in
\cite{Bazhanov:1996dr,Pronko:1999gh,Bazhanov:2001xm,Dorey:2000ma}.
The form of the operatorial equations \eqref{quadri} presented here
exactly coincide with that of the eigenvalue equations of ref. 
\cite{Tsuboi:2009ud}\footnote{%
There are also various sypersymmetric extensions of \eqref{quadri}, the
related bibliography can be found in \cite{Tsuboi:2009ud,Gromov:2010km}.}.

The most essential algebraic property of the ${\bf Q}$-operators is concisely expressed by the single 
factorization relation 
\beq
{\bf T}^+(z,\rep_n) \simeq {\bf Q}_{\{1\}}(z+\m_1')\,{\bf
 Q}_{\{2\}}(z+\m_2')\cdots {\bf Q}_{\{n\}}(z+\m_n')\label{tfact}
\eeq
where ${\bf T}^+(z,\rep_n) $ is the transfer matrix associated with
a highest weight infinite-dimensional representation (Verma module) with
the $\alg{gl}(n)$-weights $\rep_n=(\m_1,\m_2,\ldots,\m_n)$. Here
$\rep_n'=\rep_n+\rho_n$ 
denotes the weights shifted by the half sum of the positive roots,
$\rho_n$ of the algebra $\alg{sl}(n)$. Remarkably, this relation alone
allows to derive all functional relations, satisfied by various
``fusion'' transfer matrices and ${\bf Q}$-operators. As a result the
model may then be solved in an entirely algebraic fashion. 
In particular, the transfer matrices 
\beq
{\bf T}(z,\Lambda_n)\simeq{\rm det}\Vert {\bf Q}_i(z+\m'_j)\Vert_{1<i,j<n}
\eeq
associated with finite dimensional representations, are expressed as 
the determinant of a matrix constructed from Q-operators.
This approach was originally developed 
mainly for field theory models  
connected with quantized (or $q$-deformed) affine algebras
\cite{Bazhanov:1996dr,Bazhanov:1998dq,
 Bazhanov:2001xm,Bazhanov:2008yc,Kojima:2008zza}.

In \cite{Bazhanov:2010ts} we applied the same idea 
to solve the famous Heisenberg XXX spin chain corresponding to $n=2$
in \eqref{sln-ham}, which is much simpler than its $q$-deformed 
counterparts. This allowed us to observe some new algebraic 
structures which were obscured or did not manifest themselves for
field theory models. 
We found that for the XXX model the relation
\eqref{tfact} arises from very elegant factorization properties of the 
$\alg{sl}(2)$-invariant Lax operator 
\begin{equation}
\label{su2Lax}
\Lop^+(z,\Lambda_2)=z+\sum_{a=1}^3 \sigma_a \otimes {\bf J}^a=
\left( \begin{array}{cc}
\specbaz+ \SpinAux^3 \, \, &  \SpinAux^- \, \, \\
\SpinAux^+ \, \, & \specbaz-\SpinAux^3  \, \, \end{array} \right),
\end{equation}
which acts on the tensor product of a spin-$\half$ module
$\mathbb{C}^2$ and the infinite-dimensional highest weight
representation of $\alg{sl}(2)$ with an arbitrary spin $j$.  The 
$\alg{sl}(2)$ generators 
$\SpinAux^3$, $\SpinAux^\pm$ in this case are realized in 
the Holstein-Primakoff form, 
\begin{equation}
\label{su2HP}
\qquad \SpinAux^-=\oscbp
\left(\oscbp\,\oscb-2\,j  \right)\, ,\quad
\SpinAux^+=- \oscbm\, ,
\quad \SpinAux^3= j - \oscbp\,\oscbm  \, ,
\end{equation}
where $\oscbp$ and $\oscbm$ are generators of the oscillator algebra
\beq
{\mathcal H}:\qquad [\oscbm,\oscbp]=1,\qquad \osch=\oscbp\,\oscbm+\sfrac{1}{2}\ .
\eeq
Here we use $\alg{gl}(2)$ representation labels, such that $\rep_2=(j,-j)$. 
The factorization in question involves the two simpler 
``constituent'' $\Lbf$-operators
\begin{equation}
\label{su2oscLax}
\Lbf_1(z)=
\left( \begin{array}{cc}
z-\osch_1 \, \, &  ~~\oscbp_1 \, \, \\
-\oscbm_1 \, \, &~~1 \, \, \end{array} \right)
\quad {\rm and} \quad
\Lbf_2(z)=
\left( \begin{array}{cc}
1~~  \, \, & -\oscb_2  \, \, \\
\oscbp_2  \, \, & z-\osch_2 \,
\, \end{array} \right),
\end{equation}
with two sets of mutually commuting oscillators. The precise
factorization formula reads 
\begin{equation}
\label{su2fac}
\left( \begin{array}{cc}
z_1-\osch_1 \, \, &  ~~\oscbp_1 \, \, \\
-\oscb_1 \, \, &~~1 \, \, \end{array} \right)
\left( \begin{array}{cc}
1~~  \, \, & -\oscb_2  \, \, \\
\oscbp_2~~  \, \, & z_2-\osch_2 \,
\, \end{array} \right)
= 
e^{\oscb_1^+ \oscb_2^+} \left( \begin{array}{cc}
\specbaz+ \SpinAux^3 \, \, &  \SpinAux^- \, \, \\
\SpinAux^+ \, \, & \specbaz-\SpinAux^3  \, \, \end{array} \right)
\left( \begin{array}{cc}
1 \, \, & -\oscb_2 \, \,  \\
0 \, \, & 1 \, \, \end{array} \right) e^{-\oscb_1^+ \oscb_2^+}\,  ,
\end{equation}
where
\begin{equation}
\label{su2zj}
z_1=z+j+\sfrac{1}{2},\qquad z_2=z-j-\sfrac{1}{2},
\end{equation}
and the generators $\SpinAux^3$, $\SpinAux^\pm$,  
are realized in the Holstein-Primakoff form \eqref{su2HP} with $\oscbp$
and $\oscb$ replaced by $\oscbp_1$ and $\oscb_1$.
It is useful to rewrite \eqref{su2fac} in a compact form 
\beq
\Lbf_1(z_1)\,\Lbf_2(z_2)={\mathcal S}\,\Lop^+(z,\Lambda_2)\,\,\Gbb
\, \,   {\mathcal S}^{-1}\label{int:fac1},
\eeq
where
\beq\label{sg}
{\mathcal S}=e^{\oscb_1^+ \oscb_2^+},\qquad \Gbb=\left( \begin{array}{cc}
\, 1 \, \, & -\oscb_2   \\
0 \, \, & 1   \end{array} \right)\,.
\eeq 
A similar formula for the reversed order product $\Lbf_2\, \Lbf_1$ can 
also be found in \cite{Bazhanov:2010ts}.
These factorization equations 
were used to derive the $n=2$ version of \eqref{tfact} as well as the complete hierarchy of all functional relations, leading to a new algebraic solution of the XXX magnetic spin chain.  

It is interesting to note that the LHS of \eqref{su2fac} contains
two operators \eqref{su2oscLax},
which are {\em first order} polynomials in the spectral
parameter $z$, and, at the same time,
their product in the RHS is also a {\em first
 order} polynomial in $z$. This is explained by the fact that in \eqref{su2oscLax}  terms linear in
$z$ are proportional to two different degenerate
matrices whose product is equal to zero. Thus, in contrast to \eqref{su2Lax},
where the term containing $z$ is proportional to the unit matrix, the
operators \eqref{su2oscLax} start from degenerate matrices. 

By extending this key observation to the $\alg{gl}(n)$ case, we have 
completely classified all first order $\Lbf$-operators and studied their
fusion and factorization properties. In general they correspond 
to some special infinite-dimensional representations of the 
Yangian $Y(\alg{gl}(n))$ related to a direct product of
$\alg{gl}(p)\otimes {\mathcal H}^{p(n-p)}$,
involving the algebra $\alg{gl}(p)$ with $p\le n$, and a number
oscillator algebras ${\mathcal H}$.   
In particular, we find that the
$\alg{sl}(n)$ Lax operator 
\begin{equation}
\label{sunLax}
{\Lop}^+(z,{\Lambda_n})
= 
z +\sum_{a,b} e_{ab} \otimes \SpinAux_{ba}
\,
\end{equation}
for infinite-dimensional highest weight
representations of the $\alg{gl}(n)$ generators ${\bf J}_{ab}$ 
factorizes into $n$ {\it partonic} 
Lax operators $\Lbf_a(z)$ ($a=1,\ldots,n$), 
\begin{equation}
\label{boselax}
\Lbf_a(z)=\left( \begin{array}{ccccccc}
1 \, \, & \, \, & \, \, & -\oscb_{1,a} \, \,& \, \,&  \, \, & \, \,\\
\, \, &\ddots  \, \, &\, \, & \vdots  \, \,&  \, \,&\, \,& \, \, \\
\, \, &   \, \, & 1\, \, & -\oscb_{a-1,a} \, \,&  \, \,&  \, \, & \, \,\\
\oscbp_{a,1}  \, \, &   \cdots \, \, & \oscbp_{a,a-1}  \, \,& \specbaz -\osch_a \, \,& \oscbp_{a,a+1} \, \,& \cdots\, \,&\oscbp_{a,n} \, \, \\
\, \, &    \, \,&  \, \,& -\oscb_{a+1,a} \, \,& 1 \, \,&  \, \,& \, \,\\
\, \, &   \, \,&  \, \,& \vdots \, \,&   \, \,& \ddots \, \,& \, \,\\
\, \, &   \, \,&  \, \,& -\oscb_{n,a}\, \,&   \, \,& \, \,& 1\, \,\\
\end{array} \right)\,,
\end{equation}
which is the generalization of \eqref{su2oscLax} for the $\alg{gl}(n)$ case.
Each of these operators contains
$n-1$ independent oscillator pairs $(\oscb_{ba},\oscbp_{ab})$, \ 
$b=1,2,\ldots, n$,\ \  $b\neq a$,
\beq
[\oscbm_{ba},\oscbp_{ab}]=1,\qquad \osch_a=\sum_{b\neq a}\left(\oscbp_{ab}\,\oscbm_{ba}+\sfrac{1}{2}\right)\ .
\eeq
The factorization formula, generalizing \eqref{int:fac1}, now reads  
\beq\label{sunfac}
\Lbf_{1}(z+\lambda'_1)
\Lbf_{2}(z+\lambda'_2)
\cdots
\Lbf_{n}(z+\lambda'_n)
={\mathcal S}_{\Lop} \,{\Lop}^+(z\,|\,\rep_n) \,{\Gbb}_\Lop {\mathcal
 S}_\Lop^{-1}.
\eeq 
Here the shifted weights $\rep_n'=\rep_n+\rho_n$ are as in
\eqref{tfact}, $\rho_n$ is given by \eqref{rho-def} and the quantities
${\mathcal S}_{\Lop}$ and ${\Gbb}_\Lop$ are generalizations of those
in  \eqref{sg}. Eq.\eqref{sunfac} immediately implies the factorization
relation \eqref{tfact}.

Finally, note that the formula \eqref{boselax} (and its generalization
\eqref{Lcanon}) define extremely simple first order solutions of the
Yang-Baxter equation. The solutions are {\it new} and were not
previously considered in the literature. We will now proceed to derive
and precise all above statements.



\section{The Yang-Baxter Equation and Representations  of Yangians}
\label{sec:solution}
The Hamiltonian \eqref{sln-ham} with twisted
boundary conditions commutes with a large commuting family of ${\bf T}$- and
${\bf Q}$-operators. In this paper we explicitly construct these
operators via traces of certain monodromy matrices associated with
infinite-dimensional representations of the harmonic oscillator
algebra. To do this we
need to find appropriate solutions of the Yang-Baxter equation  
\begin{equation}
\label{YB-main}
\Rbf(z_1-z_2)\, \Big(\/\Lbb(z_1) \otimes 1\Big)\,
\Big(1\otimes \Lbb(z_2)\Big)= 
\Big(1\otimes \Lbb(z_2)\Big)\,\Big(\/\Lbb(z_1)\otimes 1\Big)\, \Rbf(z_1-z_2),
\end{equation}
where $\Rbf(z)$ is an $n^2 \times n^2$ matrix, 
\begin{equation}
{\Rbf}(z):\qquad 
{\mathbb C}^n \otimes {\mathbb C}^n \to {\mathbb C}^n \otimes
{\mathbb C}^n, \qquad 
{\Rbf}(z)
=z + \Pbf\, ,\label{Ryang}
\end{equation}
acting in the direct product of two $n$-dimensional spaces 
${\mathbb C}^n \otimes\,{\mathbb C}^n$. The operator ${\Pbf}$  
permutes the factors in this product. 
The operator\  ${\Lbb}(z)$ is an $n\times n$ 
matrix, acting in a single copy of the space ${\mathbb C}^n$, 
whose matrix elements are operator-valued functions of the 
variable $z$ belonging to some associative algebra ${\mathcal Y}$. 
To be more precise, the Yang-Baxter equation 
\eqref{YB-main} provides defining relations of the {\em Yangian\/} 
algebra ${\mathcal Y}= Y(\alg{gl}(n))$, introduced by Drinfeld
\cite{Drinfeld} (for a recent comprehensive review see \cite{molevbook}).

Let $L_{ij}(z)$, $i,j=1,2,\ldots,n$
denote the matrix elements of ${\Lbb}(z)$. From \eqref{YB-main}
it follows that   
\beq
(y-x)\,\big[L_{ij}(x),\,L_{k\ell}(y)\big]=
L_{kj}(x)\,L_{i\ell}(y)-L_{kj}(y)\,L_{i\ell}(x),\qquad 
i,j,k,\ell=1,2,\ldots,n.\label{lalg}
\eeq
Writing $L_{ij}(z)$ as a Laurent series with operator-valued coefficients
\beq
L_{ij}(z)=L_{ij}^{(0)}+L_{ij}^{(1)} z^{-1} +L_{ij}^{(2)}
z^{-2} + \ldots \label{laurent},
\eeq
one obtains an infinite set of commutation relations 
\beq
\big[L_{ij}^{(r)},\,L_{k\ell}^{(s)}\big]=
\sum_{a=1}^{\min(r,s)}\, \big(\,L_{kj}^{(r+s-a)}\, L_{i\ell}^{(a-1)}-
L_{kj}^{(a-1)} L_{i\ell}^{(r+s-a)}\,\big)
\label{yangcomm}
\eeq
for the elements $L_{ij}^{(r)}$, $r=0,1,2,\ldots, \infty$. Thus the
problem of solving Eq.\eqref{YB-main} reduces to the
construction of representations of the infinite-dimensional quadratic algebra \eqref{yangcomm}.
In our approach the most important role is played by 
the simplest representations, where the series
\eqref{laurent} truncates after the second term. To within a trivial change
in the normalization these
representations correspond to $\Lbf$-operators, which are first order
polynomials in the parameter $z$. All previously known
$\Lbf$-operators of this type can be brought to the form 
\beq
{\Lop}_{ij}(z)=z\,\delta_{ij}+\E_{ji}\,,\label{Leval}
\eeq
where $\E_{ij}$, \ $i,j=1,2,\dots,n$ denotes 
the standard set of generating elements of the algebra $\alg{gl}(n)$, 
\beq
\ds[\E_{ij},\E_{k\ell}] \  =
\ \ \delta_{kj}\,\E_{i\ell}\,-\delta_{i\ell}\,\E_{kj}\,.\label{gln}
\eeq
The Yang-Baxter equation for the $\Lbf$-operator \eqref{Leval} is
satisfied on the algebraic 
level by virtue of the commutation relations \eqref{gln}. Therefore 
one can choose in \eqref{Leval} arbitrary $\alg{gl}(n)$ representations 
for the generators $\E_{ij}$.  
It is obvious that the addition of a constant to the spectral parameter $z$ in 
\eqref{Leval} can be compensated by the subtraction of the same constant
from the central element 
\beq
{\mathcal C}_n=\E_{11}+\E_{22}+\ldots+\E_{nn}
\eeq
of the algebra \eqref{gln}. Therefore it is tempting to 
eliminate this spurious degree of freedom by, for instance, imposing 
the condition ${\mathcal C}_n=0$ and restricting  \eqref{gln} to the algebra 
$\alg{sl}(n)$. Here we will not do so, but will instead work with the full algebra $\alg{gl}(n)$.
This is helpful for clearly exposing the Weyl group symmetry of the problem at hand. 

We will now show that, excitingly, there exist further first order $\Lbf$-operators, different 
from \eqref{Leval}. We will present their complete classification. To begin, let us recall a simple symmetry of the Yang-Baxter equation
\eqref{YB-main}. The $R$-matrix \eqref{Ryang}
is $GL(n)$-invariant in the sense 
\beq
{\bf R}(z)=(\Gbb\otimes \Gbb)\, {\bf R}(z)\, (\Gbb\otimes \Gbb)^{-1},
\qquad \Gbb\in GL(n),\label{gl2inv}
\eeq
where $\Gbb$ is any non-degenerate $n\times n$ matrix. 
It follows that if $\Lbb(z)$ satisfies \eqref{YB-main}, then any  other 
operator of the form 
\beq
\widetilde{\Lbb}(z) =\Fbb\,\, \Lbb(z) \,\Gbb,\qquad \Fbb,\Gbb
\in GL(n),\label{Ltrans}
\eeq 
with arbitrary $\Fbb,\Gbb\in GL(n)$ satisfies again the same equation. Furthermore, the matrices $\Fbb,\Gbb$ may contain operator-valued matrix elements, as long as these commute among themselves and with all other elements of $\Lbb(z)$.

From \eqref{yangcomm} it then immediately follows that the elements 
$L_{ij}^{(0)}$ are central, i.e.~they commute among themselves and
with all $L_{ij}^{(r)}$ for $r\ge1$. Therefore, we may regard $L^{(0)}$ as a numerical $n\times n$ matrix. Applying the transformations \eqref{Ltrans}, this
matrix can always be brought to diagonal form
\beq
L^{(0)}=\mbox{diag}
\big(\underbrace{1,1,\ldots,1}_{\p\mbox{-\scriptsize{times}}},
\underbrace{0,0,\ldots,0}_{(n-\p)\mbox{-\scriptsize{times}}}\big),\qquad
\p=1,2,\ldots, n\, , 
\label{L0}
\eeq
where $\p$ is an integer $1\le \p \le n$. The number $\p$ 
coincides with the rank of the matrix $L^{(0)}$. It is invariant under the linear transformations \eqref{Ltrans}).
Evidently, if $\p=n$, the leading term in the series expansion
\eqref{laurent} is the unit matrix. This case is well studied in the existing
representation theory. In fact, the assumption that the series
\eqref{laurent} starts with the unit matrix is usually included into the
definition of the Yangian. Here we will {\it not} make this assumption,
and will consider instead the more general case with arbitrary $1\le \p \le n$.

Let us concentrate on the  
simple case when the series \eqref{laurent} truncates after the second
term, i.e.~assume that all $L_{ij}^{(r)}=0$ for $r\ge2$. It is convenient to write the
only remaining non-trivial coefficient $L^{(1)}$ as a block matrix 
\beq 
L^{(1)}=\left(
\begin{BMAT}(e){c.c}{c.c}
A_{\a\b}\ &B_{\a\dot\b}\\
C_{\dot\a\b}&D_{\dot \a \dot\b} 
\end{BMAT}\, \right)\ ,\label{L1}
\eeq
where $A,B,C$ and $D$ are operator-valued matrices of dimensions 
$\p\times \p$, $\p\times (n-\p)$, $(n-\p)\times \p$ and $(n-\p)\times (n-\p)$,
respectively. We furthermore assume that all undotted indices run over the
values $\{1,2,\ldots,\p\}$, whereas their dotted counterparts take on the values $\{\p+1,\ldots,n\}$. 
\beq
1\le\a,\b\le \p,\qquad
\p+1\le\dot\a,\dot\b\le n\ .
\eeq

Substituting \eqref{L0} and
\eqref{L1} into \eqref{yangcomm}, one immediately realizes  that the elements 
$D_{\dot\a\dot\b}$ are central, i.e.~they commute among
themselves and with all other elements of $L^{(1)}$. 
The other commutation relations read
\beq
\begin{array}{rclrclrcl}
\ds[A_{\a\b},A_{\g\e}] 
&\;=\;&\delta_{\a\e}\,A_{\g\b}-\delta_{\g\b}\,A_{\a\e}\,,\quad&
\ds[A_{\a\b},B_{\g\dot\g}]&\;=\;&-\delta_{\b\g}\,B_{\a\dot\g}\,,\quad &
\ds[A_{\a\b},C_{\dot\g\g}]&\;=\;&+\delta_{\a\g}\,C_{\dot\g\b}\,,\\[.4cm]
\ds[B_{\a\dot\b},C_{\dot\a\b}]&=&\delta_{\a\b}\,D_{\dot\a\dot\b},
&[B_{\a\dot\b},B_{\g\dot\e}]&=&0,&       
[C_{\dot\a\b},C_{\dot\g\e}]&=&0\ .
\end{array}\label{abc-comm}
\eeq
Using the remaining freedom of making 
transformations \eqref{Ltrans}, which do not affect the form of $L^{(0)}$ in 
\eqref{L0}, one
can then bring the matrix $D$ to diagonal form with zeroes and ones on
the diagonal, in similarity to \eqref{L0}. Here we are only interested in highest 
weight representations of the algebra \eqref{abc-comm}. These
representations admit a definition of the trace, as required for the
construction of transfer matrices in Section~\ref{sec:traces} below.
For this reason we only need to consider the non-degenerate case 
\footnote{%
It appears that for $\det D=0$ the algebra \eqref{abc-comm} 
does not admit a definition for a suitable trace as needed for the construction of
transfer matrices commuting with the Hamiltonian \eqref{sln-ham}.}, 
$\det D\not=0$, where the diagonal form of $D$ coincides with the
$(n-\p) \times (n-\p)$ unit matrix 
\beq
D_{\dot\a\dot\b}=\delta_{\dot\a\dot\b}, \qquad
\p+1\le\dot\a,\dot\b\le n\ . 
\eeq
The resulting algebra \eqref{abc-comm} can be realized as a direct
product of the algebra $\alg{gl}(\p)$ with $\p(n-\p)$ copies of the harmonic oscillator 
algebra:
\beq
{\mathcal A}_{n,\p}=\alg{gl}(\p)\otimes \oscalg^{\otimes
 \p(n-\p)}\,. \label{product} 
\eeq

Introduce $p(n-p)$ independent  
oscillator pairs $(\oscbm_{\dot\a\b},\oscbp_{\b\dot\a})$,\ 
where $\dot\a=p+1,\ldots,n$ and $\b=1,\ldots,p$, 
satisfying the relations
\beq
\qquad [\oscbm_{\dot\a\b},\oscbp_{\g\dot\e}]=
\delta_{\dot\a\dot\e}\,\delta_{\b\g}\,.\label{osc} 
\eeq
Furthermore, let $\E_{\a\b}$, \
$\a,\b=1,2,\ldots,p$ denote the generators of the algebra
$\alg{gl}(p)$ defined by \eqref{gln}, where $n$ is replaced by $\p$. 
The generators $\E_{\a\b}$ commute with all oscillator operators in
\eqref{osc}.  
The connection of the algebra \eqref{abc-comm} with the product
\eqref{product}  
is established by the following relations
\beq
A_{\a\b}=\oE_{\a\b}-\sum_{\dot\g=\p+1}^n \big(\oscbp_{\a\dot{\g}}\,
\oscbm_{\dot\g\b}+\sfrac{1}{2}\delta_{\a\b}\big),
\qquad  B_{\a\dot\b}=\oscbp_{\a\dot\b},
\qquad  C_{\dot\a\b}=-\oscbm_{\dot\a\b}\,,
\eeq
where the upper bar in the notation $\oE_{\a\b}$ denotes the
transposition of the indices $\a$ and $\b$, 
\beq
\oE_{{\a}{\b}}\equiv \E_{{\b\a}}.
\eeq
The corresponding  ${\bf L}$-operator can be written as a block matrix,
\beq 
\Lbf_{\{1,2,\ldots,\p\}}(z)=\left(
\begin{BMAT}(r){c.c}{c.c}
z\,\delta_{\a\b}+\oE_{\a\b}-\sum_{\dot\g=\p+1}^n\big( \oscbp_{\a\dot{\g}}\,
\oscbm_{\dot\g\b}+\sfrac{1}{2}\delta_{\a\b}\big)\ &\ \ \oscbp_{\a\dot\b}\\
-\oscbm_{\dot\a\b}&\delta_{\dot \a \dot\b} 
\end{BMAT}\, \right)\ ,\label{Lcanon}
\eeq
where the rows are labeled by the indices $\a$ or $\dot\a$ and the 
columns by $\b$ or $\dot\b$, in similarity to \eqref{L1}. Note that the
$p\times p$ matrix of the generators of $\alg{gl}(p)$, which
enters the upper left block, is transposed, i.e.~the $\a$-th row and $\b$-th column in this block contains the element
$\oE_{ab}=\E_{\b\a}$. 

The matrix \eqref{Lcanon} contains the parameter $z$ only in its first $p$
diagonal elements. By simultaneous permutations of rows
and columns in \eqref{Lcanon} one can move these $z$-containing elements to $p$ arbitrary positions on the
diagonal, labeled by a set of integers $I=\{a_1,a_2,\ldots,a_p\}$. 
We shall denote the $\Lbf$-operator obtained in this way by
${\Lbf}_I(z)$. Within this convention the operator \eqref{Lcanon}
corresponds to the set $I=\{1,2,\ldots,\p\}$, as indicated by the subscript in
the LHS of this equation. 

The ``partonic'' $\Lbf$-operator
\eqref{boselax} is a particular case of \eqref{Lcanon} with $p=1$,
while the standard $\Lbf$-operator \eqref{Leval} corresponds to $p=n$. 
We would like to stress that for $p<n$ the formula \eqref{Lcanon} yields novel solutions of the Yang-Baxter equation \eqref{YB-main}.
The only 
exception is the simple 
case $n=2$, $p=1$, where this solution was previouly known
\cite{koornwinder,Kuznetsov:1999tk,kovalsky}). Note, also that
the $n=3$ solutions can be obtained in the rational limit of 
trigonometric solutions obtained in \cite{Bazhanov:2001,Boos2010}.

The formula \eqref{Lcanon} provides an {\em evaluation homomorphism} of the
infinite-dimensional Yangian algebra \eqref{yangcomm} into the
finite-dimensional algebra \eqref{product},
\beq
Y(\alg{gl}(n))\to\alg{gl}(\p)\otimes
\oscalg^{\otimes \p(n-\p)}\label{eval-hom}\,,\qquad 1\le \p \le n\,.
\eeq
This means that for any representation of this finite-dimensional
algebra the equation \eqref{Lcanon}
automatically defines a representation of the Yangian
and a matrix solution of the Yang-Baxter
equation \eqref{YB-main}. 
Conversely, 
any first order matrix $\Lbf$-operator with a rank $\p$ leading
term $L^{(0)}$ and a non-degenerate matrix $D$ in \eqref{L1} is, up to a transformation \eqref{Ltrans}, equivalent to the {\em
 canonical} $\Lbf$-operator \eqref{Lcanon} with some particular 
representation of the algebra \eqref{product}. 
It is worth noting that the transformation \eqref{Ltrans} 
\beq
\Lbf_{\{1,2,\ldots,\p\}}(z)\rightarrow\Fbb\,\Lbf_{\{1,2,\ldots,\p\}}(z)\, \Fbb^{-1}\, , 
\eeq
where $\Fbb$ is a block diagonal matrix containing the matrices
$\Fbb_p\in GL(p)$, \ $\Fbb_{n-p}\in GL(n-p)$ on
the diagonal,  
leaves the form of \eqref{Lcanon} unchanged.

In preparation for a necessary analysis below we need to 
introduce some notation for the irreducible highest weight representations of
$\alg{gl}(n)$. The highest weight vector $v_0$ is defined by the
conditions 
\beq
\E_{a,a+1}\,v_0=0,\qquad a=1,2,\ldots,n-1.\label{hwcon}
\eeq
Here we will use $\alg{gl}(n)$-type representation labels\footnote{%
The $\alg{sl}(n)$-type weights $\mu=(\mu_1,\mu_2,\ldots,\mu_{n-1})$, 
$$
\mu_a=\m_a-\m_{a+1}, \qquad a=1,2,\ldots,n-1\,,
$$
are inconvenient from the point of view of the Weyl group
symmetry. They will not be used in this paper.}
$\rep_n=(\m_1,\m_2,\ldots,\m_n)$ where 
\beq
\E_{aa}\,v_0= \m_a\,v_0,\quad a=1,2,\ldots,n\,.
\eeq
We will denote by $\pi^+_{\rep_n}$ 
the corresponding infinite-dimensional highest weight 
representation with arbitrary
weights, and by $\pi_{\rep_n}$ the 
finite-dimensional irreducible representation with 
\beq
\m_1\ge\m_2\ge\ldots\m_n,\qquad \m_a-\m_b\in{\mathbb Z}\,.\label{domint}
\eeq


The analysis of this section extends the previous results of 
\cite{koornwinder} devoted to $n=2$ case. 
The properties of the finite dimensional
representation of the Yangian $Y(\alg{gl}(2))$ 
associated with the $\Lbf$-operator \eqref{Leval} can be
found in \cite{Tarasov1, Tarasov2, molevbook}. 
%


\section{Fusion and Factorization of \texorpdfstring{$\Lbf$}{}-operators}
\label{sec:fusion}
An essential part of our analysis in the following is based 
on some remarkable decomposition properties of the product of two
$\Lbf$-operators of the form \eqref{Lcanon}. 
The Yangian ${\mathcal Y}= Y(\alg{gl}(n))$
is a Hopf algebra, see e.g.~\cite{molevbook}. 
In particular, its co-multiplication 
\beq
{\mathcal Y}\to {\mathcal Y}\otimes 
{\mathcal Y}
\label{comul}
\eeq
is generated by the matrix product 
of two $\Lbf$-operators, corresponding to two different copies of
${\mathcal Y}$ appearing on the RHS of \eqref{comul}.

Our main observation is related to the co-product of two
operators ${\Lbf}_{I}(z)$ and ${\Lbf}_{J}(z)$, defined by \eqref{Lcanon} 
for two non-intersecting sets $I\cap J=\varnothing$,
\beq
\Lbf(z)=\Lbf_I^{[1]}(z+z_1)\,\Lbf_J^{[2]}(z+z_2),
\label{coprod}
\eeq 
where the quantities $z_{1,2}$ denote arbitrary constants. 
Let the sets $I$ and $J$ contain $\p_1$ and $\p_2$ elements, respectively. 
It is obvious that 
\begin{enumerate}[(i)]
\item
the product \eqref{coprod} is of the first order in the variable
$z$, and that 
\item
the matrix rank of the term linear in $z$ in \eqref{coprod}
is equal to $\p_1+\p_2$.
\end{enumerate}
The simple meaning of the co-multiplication is that the matrix product of two
$\Lbf$-operators, each of which satisfies by itself the Yang-Baxter equation
\eqref{YB-main}, solves this equation as well.
All solutions which possess the above properties (i) and (ii) 
were classified in the previous section.
Therefore, by using a transformation of type \eqref{gl2inv},  
the RHS of \eqref{coprod} can be 
brought to a particular case of the canonical form \eqref{Lcanon} with
$\p=\p_1+\p_2$. 
It turns out, however, that the
expressions for the matrix elements of the resulting $\Lbf$-operator 
are rather complicated and their
explicit connection to those of
\eqref{Lcanon} is far from obvious, even though these elements
satisfy the same commutation relations. In order to make this
connection more transparent we apply a suitable operatorial similarity transformation
${\mathcal S}$ to each matrix element such that it
rearranges the basis of the oscillator algebras contained in ${\mathcal
 A}_{n,\p_1}\otimes{\mathcal A}_{n,\p_2}$.  
Furthermore, the formula \eqref{coprod} contains two constants $z_1$ and
$z_2$. Obviously only their difference is an essential
parameter, whereas  the sum may be absorbed into the spectral parameter
$z$. Therefore, without loss of generality, one can set
\beq
z_1=\lambda+\sfrac{p_2}{2},\qquad 
z_2=-\sfrac{p_1}{2},
\eeq
where $\lambda$ is arbitrary. This particular parametrization is chosen to 
simplify the subsequent formulae. 

Proceeding as described above, one obtains,    
\beq
\Lbf(z)=\Lbf^{[1]}_I(z+\lambda+\sfrac{p_2}{2})
\,\Lbf^{[2]}_J(z-\sfrac{p_1}{2})={\mathcal S}\,\Big(
\Lbf_{I\cup   J}
(z)\ \Gbb\Big)\,{\mathcal S}^{-1} , \label{product2}
\eeq
where $\Gbb$ is a $z$-independent matrix, whose elements
commute among themselves and with all elements of $\Lbf_{I\cup
 J}^{\phantom{{[1]}}}(z)$. 
It should be stressed that the resulting $\Lbf$-operator $\Lbf_{I\cup
 J}^{\phantom{{[1]}}}(z)$ is only a special case of \eqref{Lcanon}, 
since it is connected to some specific realization of the algebra ${\mathcal
 A}_{n,\p_1+\p_2}$ in terms of the direct product of the two 
algebras ${\mathcal A}_{n,\p_1}\otimes{\mathcal A}_{n,\p_2}$ as
defined in \eqref{product}. Note that the RHS of \eqref{product2}
is of course a particular case of the transformation \eqref{Ltrans}
with $\Fbb\equiv1$. The explicit expression for the matrices appearing in
\eqref{product2} are presented below. Some
additional details of calculations are given in the
Appendix~\ref{app:fusion}.

By permuting rows and columns any two non-intersecting sets $I$ and
$J$ can be reduced to the case when $I=\{1,\ldots,\p_1\}$ and
$J=\{\p_1+1,\ldots,\p_1+\p_2\}$. So it is suffficient to consider this
case only. Introduce three types of indices 
\beq
\a,\b,\in I,\qquad\dot\a,\dot\b\in J,\qquad 
\ddot\a,\ddot\b\in\{\p_1+\p_2+1,\ldots,n\}\,.\label{three}
\eeq
It will be convenient to rewrite \eqref{Lcanon} as a $3\times3$ block
matrix
\begin{equation}
\Lbf_I^{[1]}(z)=\Lbf^{[1]}_{\{1,2,\ldots,\p_1\}}(z)=\left(
\begin{BMAT}(r)[0.15cm,0cm,0cm]{c.c.c}{c.c.c}
{z\,\delta_{\a\b}+\oE^{[1]}_{\a\b}-\sum_{c\not\in I} (\oscbpo_{\a c}\ 
\oscbmo_{c\b}}+\sfrac{1}{2}\delta_{\a\b}){}^{\phantom{|}}\ &\ \oscbpo_{\a\dot\b}\ &\ \oscbpo_{\a\ddot\b}\ \\
-\oscbmo_{\dot\a\b}&\delta_{\dot \a \dot\b}&0\\
-\oscbmo_{\ddot\a\b} \, \, &  0 \, \,& \, \, \delta_{\ddot\a\ddot\b} 
\end{BMAT}\, \right)\ ,\label{Lop1}
\end{equation}
where the size of the diagonal blocks is equal to $\p_1\times\p_1$,\ 
$\p_2\times\p_2$ and $(n-\p_1-\p_2)\times(n-\p_1-\p_2)$, respectively.
The superscript ``$[1]$'' indicates that the corresponding operators
belong to the ``first'' algebra in the co-multiplication
\eqref{comul}, which in the considered case is realized by the algebra
${\mathcal A}_{n,p_1}$ defined in
\eqref{product}.  Similarly, one can write $\Lbf_J^{[2]}(z)$ as
\begin{equation}
\Lbf_J^{[2]}(z)={\Lbf}_{\{p_1+1, \dots ,p_1+p_2\}}^{[2]}(z)=
\left( 
\begin{BMAT}(r)[0.15cm,0cm,0cm]{c.c.c}{c.c.c}
\delta_{\a\b} \, & -\oscbmt_{\a\dot\b}& \, \,0\\
\oscbpt_{\dot\a\b} \, \, & z\,  \delta_{\dot\a\dot\b} 
+ \oE_{\dot\a\dot\b}^{[2]} 
-\sum_{ c\not\in J}
(\oscbpt_{\dot\a c}\oscbmt_{c\dot\b}+\sfrac{1}{2}
\delta_{\dot\a\dot\b} ){}^{\phantom{|}}\, \, 
&\, \, \oscbpt_{\dot\a\ddot\b}\\
0 \, \, &  -\oscbmt_{\ddot\a\dot\b}\, \,& \, \, \delta_{\ddot\a\ddot\b} 
\end{BMAT}\, 
\right)\,,\label{Lop2}
\end{equation}
where superscript ``$[2]$'' labels operators
from the ``second'' algebra, which is the algebra ${\mathcal A}_{n,p_2}$.
By construction, all operators labeled by the superscript ``${[1]}$''
commute with 
those labeled by the superscript ``${[2]}$''. Recall also that the algebra
\eqref{product} has a direct product structure, so the generators
$\E_{\a\b}^{[1]}$ and $\E_{\dot\a\dot\b}^{[2]}$ commute with all oscillator
operators. 

With the notation introduced above the similarity transformation
${\mathcal S}$ has the form 

\beq \label{sim0}
{\mathcal S}={\mathcal S}_1\,{\mathcal S}_2\, ,
\eeq
where 
\beq\label{sim1}
{\mathcal S}_1=\exp\left(\sum_{c\in I} \sum_{ \dot c\in J}\oscbpo_{c\dot c} \oscbpt_{\dot c c}\right),
\eeq
and 
\beq\label{sim2}
{\mathcal S}_2=\exp\left(\sum_{c\in I}\sum_{\dot c\in J}\sum_{\ddot c\not \in I\cup J}\oscbpo_{c\dot c}\, \oscbpt_{\dot c\ddot c}\, \oscbmo_{\ddot c c} \right).
\eeq
The matrix $\Gbb$ has the form
\beq
\Gbb=\left( 
\begin{BMAT}(r){c.c.c}{c.c.c}
\delta_{\a\b} \, & -\oscbmt_{\a\dot\b}& \, \,0\\
0 &   \delta_{\dot\a\dot\b} 
&\, \, 0\\
0 \, \, & 0\,& \, \,\delta_{\ddot\a\ddot\b} 
\end{BMAT}\, 
\right)\,.\label{matG}
\end{equation}
The similarity transform $\mathcal{S}_1$ serves to expose the fact that the matrix entries of $\Lbf_{I\cup J}(z)$
commute with the entries of   $\Gbb$. 
The similarity transform $\mathcal{S}_2$ brings $\Lbf_{I\cup J}(z)$ to
the form \eqref{Lcanon}. 

Finally, we want to write the operator $\Lbf_{I\cup
 J}^{\phantom{{[1]}}}(z)$ in \eqref{product2} in the form
 \eqref{Lcanon} with $\p=\p_1+\p_2$. To do this we need to make the
 following identifications for the generators $\E_{ij}$,
 $i,j=1,\ldots, \p_1+\p_2$ in
 the upper diagonal block of \eqref{Lcanon}
\bea
\oJ_{\a\b}&=&\oE_{\a\b}^{[1]}-\sum_{\dot\g\in J} \oscbpo_{\a\dot{\g}}\,
\oscbmo_{\dot\g\b}+\lambda\,
\delta_{\a\b}\,,\nonumber\\[.3cm]
\oJ_{\dot\a\dot\b}&=&\oE_{\dot\a\dot\b}^{[2]}+\sum_{\g\in I}
 \oscbpo_{{\g}\dot\a}\, \oscbmo_{\dot\b\g}\,,\label{J12}\\[.3cm]
\oJ_{\a\dot\b}&=&-\oscbmo_{\a\dot\b}\,,\nonumber\\[.3cm]
\oJ_{\dot\a\b}&=&\left(\sum_{c\in I}\sum_{\dot c\in J} 
\oscbpo_{c\dot a}\,\oscbpo_{b\dot c}\,\oscbmo_{\dot c c}\right)- 
\lambda\,\oscbpo_{\dot a b}+\sum_{\dot c\in J}\oE_{\dot a\dot c}^{[2]}\, 
\oscbpo_{\dot c b}-\sum_{c\in I}\oscbpo_{\dot a c}\,\oE_{c b}^{[1]}\, ,
\nonumber
\eea
where we have used the convention \eqref{three} for numerating indices
 and $\oJ_{ij}\equiv{\mathbf J}_{ji}$. 

Furthermore, let the indices ${ \ac},\bc$ run
 over the values $1,2,\ldots,\p_1+\p_2$ and $\dot\ac,\dot\bc$ over
 the values $p_1+p_2+1,\ldots,n$. Introduce operators 
\beq\label{cosc-def}
{\bf c}_{\dot\ac\bc}=\begin{cases}\ \oscbmo_{\dot\ac\bc},\quad
\bc\in I,\\[.3cm]
\ \oscbmt_{\dot\ac\bc},\quad
\bc\in J,\\
\end{cases}
\qquad
{\bf c}^\dagger_{\ac\dot\bc}=\begin{cases}\ \oscbpo_{\ac\dot\bc},\quad
\ac\in I,\\[.3cm]
\ \oscbpt_{\ac\dot\bc},\quad
\ac\in J.\\
\end{cases}
\eeq
Then the $\Lbf$-operator $\Lbf_{I\cup   J}^{\phantom{{[1]}}}(z)$ from
\eqref{product2} can be written as  
\beq 
\Lbf_{\{1,2,\ldots,\p_1+\p_2\}}(z)=\left(
\begin{BMAT}(r){c.c}{c.c}
z\,\delta_{\ac\bc}+{\oJ}_{\ac\bc}-\sum_{\dot\gc\not\in I\cup J}
(\osccp_{\ac\dot\gc}\,
\osccm_{\dot\gc\bc}+\sfrac{1}{2}\delta_{\ac\bc}),\ &\ \ \osccp_{\ac\dot\bc}\\
-\osccm_{\dot\ac\bc}&\delta_{\dot \ac \dot\bc} 
\end{BMAT}\, \right)\ ,\label{Lcanon2}
\eeq
which has the required form as in \eqref{Lcanon}. 

The formulae \eqref{J12} 
give a homomorphism of the algebra $\alg{gl}(\p_1+\p_2)$ 
into the direct product 
\beq
\alg{gl}(\p_1+\p_2)\to
\alg{gl}(\p_1)\otimes\alg{gl}(\p_2)\otimes
\oscalg^{\otimes\,\p_1\p_2}={\mathcal B}_{\p_1,\p_2},
\label{B12}
\eeq
which for $\p_{1,2}\not=0$ has only infinite-dimensional representations (similar representations appeared in \cite{Derkachov:2008fe}).
An important feature of this map is that if one choses highest weight
representations for both algebras $\alg{gl}(p_1)$ and $\alg{gl}(p_2)$ 
then the formulae \eqref{J12} define a highest weight representation 
of $\alg{gl}(p_1+p_2)$. It is easy to check that the conditions 
\eqref{hwcon} are satisfied on the product of the corresponding highest
weight vectors $v_0^{[1]}$, $v_0^{[2]}$ and the standard Fock
vacuum for all oscillator algebras appearing in \eqref{J12}. The 
$\alg{gl}(p_1+p_2)$-weight of the resulting representation is easy to
 obtain from \eqref{J12}
\beq
\rep_{p_1+p_2}=\left(\m_1^{[1]}+\lambda,\,
\m_2^{[1]}+\lambda,\,
\ldots, \m_{p_1}^{[1]}+\lambda,\,
\m_1^{[2]},\,\m_2^{[2]},\,
\ldots,\m_{p_2}^{[2]}\right),\label{lam12}
\eeq
where $\lambda$ is an arbitrary parameter from \eqref{product2}.

The $\Lbf$-operators in the first product in \eqref{product2} have the
superscripts $[1]$ and $[2]$, which indicate that they belong to different
algebras \eqref{product} with $p=p_1$ and $p=p_2$, respectively.
By the same reason it is useful to rewrite the RHS of \eqref{product2}
supplying similar superscripts 
\beq
\eqref{product2}={\mathcal S}\,\Big(
\Lbf_{I\cup   J}^{{{[1']}}}
(z)\ \Gbb^{[2']}\Big)\,{\mathcal S}^{-1} , \label{product3}
\eeq
where the superscript $[1']$ indicates the algebra \eqref{B12} and the
superscript $[2']$ indicates the product of oscillator algebras
${\mathcal H}^{\otimes p_1p_2}$.  Note that the matrix $\Gbb$ could
be considered as a $z$-independent $\Lbf$-operator, also satisfying the
Yang-Baxter equation \eqref{YB-main}. In view of this Eq.\eqref{product3} 
also describes the co-multiplication of two representations of the Yangian.  

Consider now some particular consequences of formula \eqref{product2}.  
Using it iteratively with $p_1=1$ and taking into account
\eqref{lam12} one can obtain an arbitrary product of the elementary 
$\Lbf$-operators \eqref{boselax}. Let $I=(a_1,\ldots,a_p)$ be an ordered
integer set, $1\le a_1<\ldots<a_p\le n$, and ${\mathcal
 L}^+_I(z\,|\,\rep_p)$ a specialization of the $\Lbf$-operator
\eqref{Lcanon} to the infinite dimensional highest weight
representation $\pi^+_{\rep_p}$ of the algebra $\alg{gl}(p)$ 
\beq
{\Lbf}^+_I(z\,|\,\rep_p)=\pi^+_{\rep_p}\big[\Lbf_I(z)\big],\qquad
\rep_p=
(\m_1,\m_2,\ldots,\m_p)\,.\label{Lp-plus}
\eeq
Define also the shifted weights (cf. \eqref{rho-def})
\beq
\rep'_p=
(\m'_1,\m'_2,\ldots,\m'_p)\,,\qquad \m'_j=\m_j+\sfrac{p-2j+1}{2},\qquad
j=1,\ldots,p\,.\label{shifted}
\eeq 
Then it follows that from \eqref{product2}
\beq
\Lbf_{a_1}(z+\lambda'_1)
\Lbf_{a_2}(z+\lambda'_2)
\cdots
\Lbf_{a_p}(z+\lambda'_p)
={\mathcal S}_I \,{\Lbf}^+_I(z\,|\,\rep_p) \,{\Gbb}_I {\mathcal
 S}_I^{-1}\label{factor1}
\eeq 
where the matrices ${\mathcal S}_I$ and $\Gbb_I$ are products of the
expressions of the type \eqref{sim0} and \eqref{matG} arising from the
repeated use of the formula \eqref{product2}. In the particular case 
$p=n$ the last formula provides the factorization for the
$\Lbf$-operator \eqref{Leval},
\beq
\Lop^+(z\,|\,\rep_n)=\pi^+_{\rep_n}\big[\Lop(z)\big],
\eeq
evaluated for the infinite-dimensional  
highest weight representation $\pi^+_{\rep_n}$ in the auxiliary space,
\beq\label{factor2}
\Lbf_{1}(z+\lambda'_1)
\Lbf_{2}(z+\lambda'_2)
\cdots
\Lbf_{n}(z+\lambda'_n)
={\mathcal S}_{\Lop} \,{\Lop}^+(z\,|\,\rep_n) \,{\Gbb}_\Lop {\mathcal
 S}_\Lop^{-1}.
\eeq 
An independent proof of this fact is given 
in the appendix \ref{app:factorization}.

\section{Construction of the \texorpdfstring{$\bf Q$}{}-operators}
\label{sec:traces}
The purpose of this section is to define the $\bf T$- and $\bf
Q$-operators. They have to commute with the Hamiltonian \eqref{sln-ham} of
the twisted compact $\alg{gl}(n)$-spin chain of length  
$L$.  These operators act on the quantum space ${\mathcal V}$, 
which is an $L$-fold tensor product of the fundamental representations 
of the algebra $\mathfrak{gl}(n)$,
\begin{equation}\label{quantumspace}
{\mathcal V}=\underbrace{
{\mathbb C}^n\otimes {\mathbb C}^n\otimes \cdots \otimes  
{\mathbb C}^n}_{L-\mbox{\scriptsize{times}}}
\ .
\end{equation}
As before, solutions of the Yang-Baxter equation 
\eqref{YB-main} are considered as $n$ by $n$ matrices, acting in the
quantum space of a single spin. Their matrix elements are operators   
in some representation space $V$ of the Yangian algebra \eqref{lalg}.
This representation space will be called here the auxiliary space. 
For each solution of \eqref{YB-main} one can define a transfer matrix, 
\begin{equation}
{\mathbb T}_{V}(z)=\mbox{Tr}_{V} \Big\{{\mathbb D}\, {\Lbb}(z)\otimes
{\Lbb}(z)\otimes\cdots \otimes{\Lbb}(z)\Big\},
\label{T-gen}
\end{equation}
where the tensor product is taken with respect to the quantum spaces
${\mathbb C}^n$, while the operator product and the trace is taken
with respect to the auxiliary space $V$. The quantity 
${\mathbb D}$ is a ``boundary twist'' operator acting only in
auxiliary space, i.e.~it acts trivially in the quantum space.
This boundary operator is completely determined by the requirement of
commutativity of the 
transfer matrix \eqref{T-gen} with the Hamiltonian \eqref{sln-ham},
which leads to the following conditions 
\begin{equation}\label{D-def}
{\mathbb D}\,\big({\Lbb}(z)\big)_{ab} \,{\mathbb D}^{-1}=
e^{i\,\left(\Phi_b-\Phi_a\right)}\, \big( {\Lbb}(z)\big)_{ab}\,, 
\qquad a,b=1,\ldots,n\,,
\end{equation}
where the fields $\Phi_a$ are the same as in \eqref{bcH}. Note that
these fields enter \eqref{D-def} only through their differences. It 
is convenient to set 
\begin{equation}
\sum_{a=1}^n \Phi_a=0\, .
\end{equation}
Solving \eqref{D-def} for the general $\Lbf$-operator \eqref{Lcanon} with an
arbitrary set $I=\{a_1,a_2,\ldots,a_p\}$, one obtains  
\beq\label{D-exp}
\Dbf_I=\exp\Big\{i\sum_{a\in I} \Phi_a \E_{aa}-
i\sum_{a\in I}\sum_{\dot\b\not\in I} (\Phi_a-\Phi_{\dot\b}) \,\oscbp_{a\dot\b}
\oscbm_{\dot\b\a}\Big\}\,.
\eeq

In the following we will use some important properties of the 
trace over the Fock representations of the oscillator algebra 
\beq
[\,\oscb,\oscb^{\dagger}\,]=1,\quad [\osch,\oscb]=-\oscb,
\quad [\osch,\oscb^\dagger\,]=\oscb^\dagger,\quad
\osch=\oscbp\,\oscb+\sfrac{1}{2}\ .\label{osc3}
\eeq
This algebra has two Fock representations, 
\beq
{\mathcal F}_+:\qquad \oscb\,|0\rangle=0, \qquad
|k+1\rangle=\oscb^\dagger\,|k\rangle, 
\eeq
and 
\beq
{\mathcal F}_-:\qquad \oscb^\dagger\,|0\rangle=0, \qquad
|k+1\rangle=\oscb \,|k\rangle, 
\eeq
spanned on the vectors
$|k\rangle$,\ \ $k=0,1,2,\ldots \infty$.
These representations can be obtained from each other via a
simple automorphism of \eqref{osc3},
\beq
\oscb \to -\oscb^\dagger,\qquad \oscb^\dagger \to \oscb,\qquad 
\osch\to-\osch\,.
\eeq
Let $P(\oscb,\oscbp)$ be an arbitrary polynomial of the
operators $\oscb$ and $\oscbp$. 
Below it will be convenient to use a {\em normalized trace} over the
representations ${\mathcal F}_\pm$,
\beq
\ntr_{\mathcal F} \Big\{e^{i\Phi \osch}
P(\oscb,\oscbp)\Big\}\ \ \mathop{=}^{\mbox{\small 
def}}\ \ \frac { \mbox{Tr}_{\mathcal F} \Big\{e^{i\Phi\osch}
P(\oscb,\oscbp)\Big\}_{\phantom{|}}}{\mbox{Tr}_{\mathcal F} 
\Big\{e^{i\Phi\osch}\Big\}^{\phantom{|}}}\ ,\qquad {\mathcal F}={\mathcal
 F}_\pm\, ,
\label{norm-tr}
\eeq
where ${\mathcal F}$ is either ${\mathcal F}_+$ or ${\mathcal F}_-$,
and $\mbox{Tr}_{\mathcal F}$ denotes the standard trace.  An
important feature of the normalized trace \eqref{norm-tr} is that it is
completely determined by the commutation relations \eqref{osc3}
and the cyclic property of the trace. It is therefore independent of a particular 
choice of representation as long as the traces in the RHS
of \eqref{norm-tr} converge.  Alternatively, one can reproduce the same
result by using explicit expressions for the matrix elements of the oscillator operators in 
\eqref{norm-tr}. Then the trace over ${\mathcal F}_+$ converges when
$\mbox{Im}\,\Phi>0$ and the trace over ${\mathcal F}_-$ when
$\mbox{Im}\,\Phi<0$. Both ways of calculation lead to the same analytic
expression for the normalized trace. Thus it is not necessary to
specify which of the two representations ${\mathcal F}_\pm$ is used.

We are now ready to define various transfer matrices all commuting
with the Hamiltonian \eqref{sln-ham}. Consider the most general
$\Lbf$-operator \eqref{Lcanon} with an arbitrary set
$I=\{a_1,a_2,\ldots,a_p\}$, where $p=1,2,\ldots,n$. Recall that the
matrix elements of \eqref{Lcanon} belong to the direct product
\eqref{product} of the algebra $\alg{gl}(\p)$ and of $\p(n-\p)$
oscillator algebras. Choose a finite-dimensional representation
$\pi_{\Lambda_{p}}^{\phantom{+}}$ with the highest weight
${\Lambda_{p}}$ for the $\alg{gl}(p)$-factor of this
product.  Then substituting \eqref{Lcanon} and \eqref{D-exp} into
\eqref{T-gen} one can define rather general transfer matrices
\beq
{\Xbf}_I(z,\Lambda_{p})=
e^{i z\,(\,\sum_{a\in I}\Phi_a)
}\ \mbox{Tr}_{\pi_{\Lambda_{p}}}
\ntr_{{\mathcal F}^{\p(n-\p)}} 
\big\{\/{\bf M}_I(z)\big\}\,,\label{Z-def}
\eeq
where ${\bf M}_I(z)$ is the corresponding monodromy matrix, 
\beq
\qquad {\bf M}_I(z)=
{\Dbf}_I\, {\mathbf L}_I(z)\otimes
{\mathbf L}_I(z)\otimes\cdots \otimes{\mathbf L}_I(z)\,.\label{M-def}
\eeq
Here $\ntr_{{\mathcal F}^{\p(n-\p)}}$ denotes the normalized trace
\eqref{norm-tr} for all involved oscillator algebras\footnote{%
It is
easy to check that all possible expressions under the trace in
\eqref{Z-def} for each oscillator algebra are exactly as in the LHS of
\eqref{norm-tr} for some polynomial $P$ and some 
value of $\Phi$. Thus the definition \eqref{norm-tr} is sufficient to
calculate all oscillator traces in \eqref{Z-def}.}, 
while $\mbox{Tr}_{\pi_{\Lambda_p}}$
denotes the standard trace over the representation $\pi_{\Lambda_\p}$
of $\alg{gl}(\p)$.  The exponential scalar factor in front of the
trace is introduced for later convenience. 

Similarly one can define a related quantity where the
$\alg{gl}(p)$-trace is taken over an infinite-dimensional highest
weight representation $\pi_{\Lambda_{p}}^{{+}}$,
\beq
{\Xbf}_I^+(z,\Lambda_{p})=
e^{i z\,(\,\sum_{a\in I}\Phi_a)
}\ \mbox{Tr}_{\pi_{\Lambda_{p}}^+}
\ntr_{{{\mathcal F}^{\p(n-\p)}}^{\phantom{|}}} 
\big\{\/{\bf M}_I(z)\big\}\,,\label{Zp-def}
\eeq
while the rest of the expression remains the same as in
\eqref{Z-def}. Note that in the case of \eqref{Z-def} the weights
$\Lambda_p=(\lambda_1,\lambda_2,\ldots,\lambda_{p})$ satisfy the 
conditions \eqref{domint}. In contradistinction, in
\eqref{Zp-def} these weights are arbitrary.   

In the limiting case $p=n$, the general $\Lbf$-operator \eqref{Lcanon}
simplifies to \eqref{Leval}, while the expression \eqref{D-exp} simplifies to 
\beq
\Dbf=\Dbf_{\{1,2,\ldots,n\}}
=\exp\Big\{i\sum_{a=1}^n \Phi_a \E_{aa}\Big\}\label{sln-exp}\,.
\eeq
In this case the definition \eqref{Z-def} reduces to that for the standard
${\bf T}$-operator
\beq
{\bf T}_{\Lambda_n}(z)\equiv{\Xbf}_{\{1,2,\ldots,n\}}(z,{\Lambda_n})=
{\rm Tr}_{\pi_{\Lambda_n}}\Big\{\Dbf\,{\Lop}(z)\otimes{\Lop}(z)
\cdots \otimes{\Lop}(z)\Big\}\, ,
\label{t-def}
\eeq
associated with the finite-dimensional representation
${\pi_{\Lambda_n}}$ of the algebra $\alg{gl}(n)$ in the auxiliary
space. Here $\Lop(z)$ denotes
the ${\bf L}$-operator \eqref{Leval}.
Likewise, the formula \eqref{Zp-def} reduces to the ${\bf T}$-operator
\beq
{\bf T}^+_{\Lambda_n}(z)={\Xbf}^+_{\{1,2,\ldots,n\}}(z,{\Lambda_n}) 
\label{tplus}
\eeq
associated with the infinite-dimensional representation
$\pi_{\Lambda_n}^+$. The above two ${\bf T}$-operators are connected
due the Bernstein Gel'fand Gel'fand (BGG) resolution of the finite
dimensional modules \cite{BGG}. The BGG result allows one to express
finite dimensional highest weight modules in terms of an alternating sum of 
infinite dimensional highest weight modules. This implies that 
the $\Top$-operator \eqref{t-def} for a finite dimensional module 
can be written in terms of \eqref{tplus} as
\beq
{\bf T}_{\Lambda_n}(z)=\sum_{\sigma\in S_n}(-1)^{l(\sigma)}\, 
{\bf T}^+_{\sigma(\Lambda_n+\rho_n)-\rho_n}(z),
\label{bgg1}
\eeq
where $\rho_n$ is a constant $n$-component vector
\beq
\rho_n=\Big(\sfrac{n-1}{2},\sfrac{n-3}{2},\ldots,\sfrac{1-n}{2}\Big),
\label{rho-def}
\eeq
which coincides with the half sum of the positive roots of the algebra
$\alg{sl}(n)$. The summation in \eqref{bgg1} 
is taken over all permutations of $n$
elements, $\sigma\in S_n$,  and $l(\sigma)$ is the parity of the permutation
$\sigma$. The relation \eqref{bgg1} and its connection to the BGG
resolution were first obtained in \cite{Bazhanov:2001xm} in the context of
$U_q(\widehat{sl}(3))$, while the $n=2$ case was previously considered in
\cite{Bazhanov:1996dr,Bazhanov:1998dq}. 

Similarly, for \eqref{Z-def} one has
\beq
{\Xbf}_I(z,\Lambda_p)=\sum_{\sigma\in S_p}(-1)^{l(\sigma)}\, 
{\Xbf}_I^+(z,\sigma(\Lambda_p+\rho_p)-\rho_p),
\label{bgg2}
\eeq
where $\rho_p$ is a $p$-component vector defined as in \eqref{rho-def}
with $n$ replaced by $p$.

Another limiting case of \eqref{Z-def} corresponds to the representation
$\pi^{\phantom{+}}_{\Lambda_\p}$ turning into the trivial
one-dimensional representation of $\alg{gl}(p)$ with weight 
$\Lambda=(0,0,\ldots,0)$. As we shall see below the
resulting operators 
\beq
{\bf Q}_I(z)={\Xbf}_I(z,(0))
\label{QI-def}
\eeq
are actually the ${\bf Q}$-operators, whose eigenvalues appear in the
nested Bethe Ansatz equations.  Let us enumerate these ${\bf Q}$-operators. 
It is convenient to start formally from the exceptional case $p=0$,
corresponding to an empty set $I=\varnothing$. By definition we set 
\beq
{\bf Q}_\varnothing(z)\equiv 1.\label{Q0}
\eeq
For the next level $p=1$ there are obviously
$n$ sets $I$ consisting of just one
element $I=\{a\}$, \ $a=1,2,\ldots,n$. The general $\Lbf$-operator
\eqref{Lcanon} in this case takes the simple form \eqref{boselax}
and the twist operator \eqref{D-exp} simplifies to 
\beq \Dbf_{a}\equiv\Dbf_{\{a\}}=
\exp\Big\{-i\sum_{\dot\g\not\in I} (\Phi_a-\Phi_{\dot\g}) \,\oscbp_{a\dot\g}
\oscbm_{\dot\g\a}\Big\}\,,
\qquad a=1,2\ldots,n.\label{D-parton} \eeq
In this way one obtains from \eqref{Z-def}
\beq
{\bf Q}_a(z)\equiv{\bf Q}_{\{a\}}(z)={\Xbf}_{\{a\}}(z,(0))=
e^{i z\,\Phi_a
}\ \ntr_{{\mathcal F}^{(n-1)}} 
\Big\{{\Dbf}_a\, {\mathbf L}_a(z)\otimes
{\mathbf L}_a(z)\otimes\cdots \otimes{\mathbf L}_a(z)\Big\}
\,,\label{Qa-def}
\eeq
where $a=1,2\ldots,n$ and ${\mathbf L}_a(z)$ is given by \eqref{boselax}.

More generally, for the level $p$ there are $\binom{n}{p}$ increasing
integer sets $I=\{a_1,\ldots,a_p\}\subseteq \{1,2,\ldots,n\}$, which 
numerate the ${\bf Q}$-operators \eqref{QI-def}. For the highest
level $p=n$ the definitions \eqref{QI-def} and \eqref{Z-def}
immediately lead to the result 
\beq
{\bf Q}_{\{1,2,\ldots,n\}}(z)=z^L,\label{Qn}
\eeq
where $L$ is the length of the chain. Altogether there are $2^n$
different ${\bf Q}$-operators\footnote{%
 The ${\bf Q}$-operators can be conveniently associated with nodes of a hyercubical Hasse diagram. They are labelled by their index sets
 $I=\{a_1,a_2,\ldots,a_p\}$, ordered by inclusion. See next section, and
 Appendix~\ref{App:Hasse}.}
 , including \eqref{Q0} and \eqref{Qn}.
These operators form in conjunction with all ${\bf T}$ and ${\bf X}$-operators a commuting family and therefore can be simultaneously
diagonalized. It is easy to see that their eigenvalues have the form
\beq\label{Q-eigen}
{\rm Q}_I(z)=e^{i z\,(\,\sum_{a\in I}\Phi_a)} \, \prod_{k=1}^{m_I}
(z-z^I_k),\qquad m_I=\sum_{a\in I} m_a\,,
\eeq
where, for each eigenstate, the numbers 
$m_a$ are the conserved occupation numbers,
\beq
m_1+m_2+\cdots+m_n=L\,,
\eeq
defined in the introduction.

We would like to stress, 
that in general the operators ${\bf X}_I(z,\rep_p)$, defined in 
\eqref{Z-def}, involve the trace over
a representation of the Lie algebra $\alg{gl}(p)$ and the trace over a
number of Fock
representations of the oscillator algebra, whereas the ${\bf
 T}$-operators involve only the Lie algebra trace and the ${\bf
 Q}$-operator only the oscillator traces. This is why we denoted the 
``hybrid'' operators \eqref{Z-def} by a distinct symbol ${\bf X}$
(thereby continuing a steady tradition of the field which already have
${\rm Y}$-systems, ${\rm T}$-systems, ${\rm Q}$-systems, etc. to occupy all letters of
the alphabet).

\section{Functional Relations}
\label{sec:func}

The results of Section~\ref{sec:fusion} imply various functional relations 
for the ${\bf Q}$-operators. To derive them we need to use some
additional properties of the twist operators \eqref{D-exp} which are 
not immediately obvious from their definition \eqref{D-def}. 
Let $\Dbf_I$ and $\Dbf_J$ be the operators \eqref{D-exp},
corresponding to $\Lbf^{[1]}_I(z)$ and $\Lbf^{[2]}_J(z)$ from the LHS
of \eqref{product2}. By explicit calculation one can check that 
the product of these operators commutes
with the similarity
transformation ${\mathcal S}$ defined in \eqref{sim0},
\beq
\big[\Dbf_I\,\Dbf_J\,,\, {\mathcal S}\big]=0\,.\label{D1}
\eeq
Moreover this product can be rewritten in the form
\beq
\Dbf_I\,\Dbf_J=\Dbf_{I\cup J}\,\Dbf_{\mathbb G}\, ,
\label{D2}
\eeq
where $\Dbf_{I\cup J}$ and  $\Dbf_{\mathbb G}$ are the twist
 operators obtained from \eqref{D-def} for the operator 
$\Lbf_{I\cup   J}(z)$ and 
the $z$-independent $\Lbf$-operator $\Gbb$ from the RHS of
\eqref{product2}. Again the relation \eqref{D2} 
is verified by direct calculation, where one needs to take into
account the explicit form of \eqref{matG}, \eqref{J12},
\eqref{cosc-def} and \eqref{Lcanon2}.

Next, define a scalar factor, {\it cf.}~\eqref{Delta},
\beq\label{vandermond}
\Delta_I(\Phi)=\Delta_{\{a_1,a_2,\ldots,a_p\}}(\Phi)=
\prod_{1\le i<j\le p} 2i\,\sin\bigg(\frac{\Phi_{a_i}-\Phi_{a_j}}{2}\bigg),
\eeq
which depends on the set $I$ and the fields
$\Phi_1,\Phi_2,\ldots,\Phi_n$. Combining \eqref{product2} with the
definition \eqref{Zp-def} and taking into account \eqref{D1} and
\eqref{D2} one obtains
\begin{equation}
\Delta_I\,
\Xbf^+_{I}(z+\sfrac{p_2}{2},\Lambda_{p_1})\ 
\Delta_{ J}\,\Xbf^+_{J}(z+\lambda-\sfrac{p_1}{2},\Lambda_{p_2})=
{\Delta_{{I}\cup{J}}}\,
\Xbf^+_{{I}\cup{J}}(z,\Lambda_{p_1+p_2}).
\label{func-main}
\end{equation}   
There are two nontrivial steps in the derivation of the last formula which
require explanations. First, the simple transfer matrix 
\beq
{\bf T}_{\Gbb}=\ntr_{{\mathcal F}^{p_1p_2}}\Big\{\Dbb_{\Gbb}
\Gbb\otimes\Gbb\otimes\cdots\otimes\Gbb\Big\}=1
\eeq
that arises in the calculations is equal to the identity
operator. Second, the scalar factors in \eqref{func-main} arise due to
the difference in the definition of the trace over the oscillator
algebras (normalized trace \eqref{norm-tr}) and over the
representation of $\alg{gl}(p)$ (standard trace). From \eqref{B12} it
is clear that $p_1p_2$ oscillator pairs have to be ``redistributed'' to
support the Holstein-Primakoff realization of the
infinite-dimensional representation $\pi^+_{\rep_{p_1+p_2}}$ of the
algebra $\alg{gl}(p_1+p_2)$.

A particular simple case of \eqref{func-main} arises when $p_2=1$ and 
$J=\{a_{p+1}\}$,
\begin{equation}
\Delta_I\,
\Xbf^+_{I}(z+\sfrac{1}{2},\Lambda_{p})\ 
{\bf Q}_{a_{p+1}}(z+\lambda_{p+1}-\sfrac{p}{2})=
{\Delta_{{I}\cup{a_{p+1}}}}\,
\Xbf^+_{{I}\cup{a_{p+1}}}(z,\Lambda_{p+1}),
\end{equation}   
where by the definition \eqref{Qa-def} one has
${\bf Q}_a(z)\equiv{\Xbf}_{\{a\}}(z,(0))$. Iterating the last formula
one obtains 
\beq
\Delta_I\ \Xbf_I^+(z,\rep_p) = {\bf Q}_{a_1}(z+\m_1')\,{\bf
 Q}_{a_2}(z+\m_2')\cdots {\bf Q}_{a_p}(z+\m_p'),
\eeq
where the notation here is the same as in \eqref{shifted} and
\eqref{factor1}. Next, applying \eqref{bgg2} one gets 
\beq\label{Xasdet}
\Delta_I\ \Xbf_I(z,\rep_p) = {\det} \Vert\,
{\bf Q}_{a_i}(z+\m_j')\,\Vert_{1\le i,j \le p}, 
\eeq
and setting $\rep_p=(0)$ one finally arrives at 
\beq\label{Qasdet}
\Delta_I\ {\bf Q}_I(z) = {\det} \Vert\,
{\bf Q}_{a_i}(z-j+\sfrac{p+1}{2})\,\Vert_{1\le i,j \le p}\, . 
\eeq
Note also that in a particular case $I=\{1,2,\ldots,n\}$ the formula 
\eqref{Xasdet}  
leads to the determinant expression \cite{Bazhanov:1996dr,Krichever:1996qd,
Bazhanov:1998dq,
 Bazhanov:2001xm,Kojima:2008zza} 
for transfer matrix \eqref{t-def},
\beq
\Delta_{\{1,2,\ldots,n\}}\ {\bf T}_{\Lambda_n}(z)=
{\det} \Vert\,
{\bf Q}_{i}(z+\m_j')\,\Vert_{1\le i,j \le n}\ ,
\eeq
where $\Delta_{\{\ldots\}}$ is defined in \eqref{vandermond} and
$\lambda_j'$ in \eqref{shifted}.


As previously mentioned the $2^n$ different operators
$\Qop_I$ can be assigned to the nodes of a hypercubic Hasse diagram.  We will
now show that four ${\bf Q}$-operators belonging to the same
quadrilateral as in Fig.~\ref{Hasse0} satisfy a remarkably simple
functional equation, which can be identified with the famous Hirota
equation from the theory of classical discrete evolution equation. 
Define a matrix
\beq
\label{matyrixofQ}
M_{ij}\equiv \Qop_{a_i}(z-j+\sfrac{p+1}{2})\,,
\quad i,j\in \{0,\ldots,p+1\}\, .
\eeq
where $\{a_0,a_1,\ldots,a_p,a_{p+1}\}$ in an increasing sequence of
 $p+2$ integers which contains the subsequence
 $I=\{a_1,\ldots,a_p\}$. Denote $a\equiv a_0$ and $b\equiv a_{p+1}$.
\begin{figure}[t]
\includegraphics[scale=0.15]{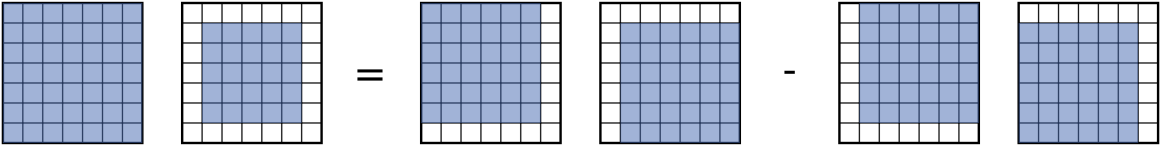}
\caption{Jacobi determinant formula.}
\label{Fig:Jacobi}
\end{figure}
Let us now use
Jacobi's formula for determinants (see Fig.~\ref{Fig:Jacobi}) for the 
matrix \eqref{matyrixofQ}. Applying \eqref{Qasdet} for the
subdeterminants one obtains the following
operatorial functional relation\footnote{%
More general relations involving operators \eqref{Xasdet} can be
obtained in the same way by replacing the arguments of the ${\bf
 Q}$-operators in \eqref{matyrixofQ} 
with arbitrary constants $z_j$, $j=0,\ldots,p+1$.}
\beq
\label{FormJacobi}
\Delta_{\{a,b\}}\,\Qop_{I\cup a \cup
b}(\specbaz)\,\Qop_{I }(\specbaz)= \Qop_{I \cup
a}(\specbaz-\half)\,\Qop_{I \cup b}(\specbaz+\half)- \Qop_{I
\cup b}(\specbaz-\half)\,\Qop_{I \cup a}(\specbaz+\half)\, ,
\eeq
which was already stated in the introduction as \eqref{quadri}, and
where $\Delta_{a,b}$ is given in \eqref{Delta}. 
Since all $\Qop$-operators commute with each other the same relation
\eqref{FormJacobi} holds also for the corresponding eigenvalues. We
will use it to derive Bethe equations in the next section. 



\section{Bethe Equations}

The connection between the Hirota equations \eqref{FormJacobi} and the
Bethe Ansatz equations is well understood 
\cite{Bazhanov:1996dr,Pronko:1999gh,Bazhanov:2001xm,Dorey:2000ma,
  Kazakov:2007fy,Tsuboi:2009ud}.    
Consider the equations \eqref{FormJacobi} at the eigenvalue level. The
reader might find it useful to look at the examples of Hasse diagrams in the introduction and
in appendix \ref{App:Hasse} when following through the following
derivation.  Let us denote the zeroes of ${\rm Q}_{I \cup a}(\specbaz)$ by
$\specbaz_k^{I \cup a}$.  Taking $\specbaz+\half=\specbaz_k^{I \cup
a}$ and $\specbaz-\half=\specbaz_k^{I \cup a}$, equation
\eqref{FormJacobi} reads
\beq
{\rm Q}_{I\cup a \cup b}(\specbaz_k^{I \cup a}-\half)\,{\rm Q}_{I }(\specbaz_k^{I \cup a}-\half)\sim 
{\rm Q}_{I \cup a}(\specbaz_k^{I \cup a}-1)\,{\rm Q}_{I \cup b}(\specbaz_k^{I \cup b})\,,
\eeq
\beq
{\rm Q}_{I\cup a \cup b}(\specbaz_k^{I \cup a}+\half)\,{\rm Q}_{I }(\specbaz_k^{I \cup a}+\half)\sim 
- {\rm Q}_{I \cup a}(\specbaz_k^{I \cup a}+1)\,{\rm Q}_{I \cup b}(\specbaz_k^{I \cup b})\,,
\eeq
respectively, where $a,b\not \in I$.
Taking the ratio of these two equations above one obtains
\beq
\label{formalBETHE}
-1=\frac{{\rm Q}_{I }(\specbaz_k^{I \cup a}-\half)}{{\rm Q}_{I }(\specbaz_k^{I \cup a}+\half)}\,
\frac{ {\rm Q}_{I \cup a}(\specbaz_k^{I \cup a}+1)}{ {\rm Q}_{I \cup a}(\specbaz_k^{I \cup a}-1)}\,
\frac{{\rm Q}_{I\cup a \cup b}(\specbaz_k^{I \cup a}-\half)}{{\rm Q}_{I\cup a \cup b}(\specbaz_k^{I \cup a}+\half)}\,.
\eeq
Here $I$ can also be the empty set. In this case we can remove ${\rm Q}_{\varnothing}(z)$ from
the equation using \eqref{Q0} . The number of elements in $I\cup a\cup b$ cannot
exceed $n$, therefore $I$ contains at most $n-2$ elements. 
Thus one obtains $n-1$ different relations of the type
\eqref{formalBETHE}, with various cardinalities of the set $I$. This
exactly matches the number of levels of nested Bethe equations for the
$\mathfrak{gl}(n)$-spin chain. 

Let us take any sequence $(a_1,\ldots,a_n)$ of elements of the set
$\{1,\ldots,n\}$. We construct a sequence of ascending sets
$\varnothing=I_0\subset I_1\subset\ldots\subset I_n=\{1,\ldots,n\}$ such
that\footnote{
This construction can easily be
obtained by choosing a path on the Hasse diagram leading from
$\varnothing$ to $\{1,\ldots, n\}$. See Appendix \ref{App:Hasse} for
more details.}
$I_i=I_{i-1}\cup a_i$. Then for each $I_i$, $i=1,\ldots, n-1$, we can rewrite
\eqref{formalBETHE} as
\begin{equation}\label{formalBETHE2}
-1=\frac{{\rm Q}_{I _{i-1}}(\specbaz_k^{I_i}-\half)}{{\rm Q}_{I_{i-1} }(\specbaz_k^{I _i}+\half)}\,
\frac{{\rm Q}_{I_i}(\specbaz_k^{I_i}+1)}{{\rm Q}_{I_i}(\specbaz_k^{I_i}-1)}\,
\frac{{\rm Q}_{I_{i+1}}(\specbaz_k^{I_i}-\half)}{{\rm Q}_{I_{i+1}}(\specbaz_k^{I_i}+\half)}\,.
\end{equation}
We will call an equation with $z^{I_i}$ roots the $i$-th level
equation. One need to use \eqref{Q0} for the lowest level, $i=1$,  
and \eqref{Qn} for the highest
level equation with $i=n-1$.

It is not difficult to see that the system of functional equations
\eqref{formalBETHE} already corresponds to the nested system of Bethe
equations of the $\alg{gl}(n)$ compact magnets. To recover their
traditional form, we merely need to substitute the eigenvalue formula
\eqref{Q-eigen} into \eqref{formalBETHE2}. The crucial point of our
approach is that the analytic structure of \eqref{Q-eigen} 
is rigorously derived without any assumptions. 
It immediately {\it follows} from the
explicit construction of the $\Qop$-operators in Section 4. 
Then one can rewrite \eqref{formalBETHE2} as 
\begin{equation}\label{BE1}
e^{i(\Phi_{a_{2}}-\Phi_{a_1})}=\prod_{k\neq l}\frac{z_l^{I_1}-z_k^{I_1}+1}{z_l^{I_1}-z_k^{I_1}-1}\prod_k\frac{z_l^{I_1}-z^{I_{2}}-\sfrac{1}{2}}{z_k^{I_1}-z^{I_{2}}+\sfrac{1}{2}}\,,
\end{equation}
for the lowest level,
\begin{equation}
e^{i(\Phi_{a_{i+1}}-\Phi_{a_i})}=\prod_k\frac{z_l^{I_i}-z_k^{I_{i-1}}-\sfrac{1}{2}}{z_l^{I_i}-z_k^{I_{i-1}}+\sfrac{1}{2}}\prod_{k\neq l}\frac{z_l^{I_i}-z_k^{I_i}+1}{z_l^{I_i}-z_k^{I_i}-1}\prod_k\frac{z_l^{I_i}-z^{I_{i+1}}-\sfrac{1}{2}}{z_k^{I_i}-z^{I_{i+1}}+\sfrac{1}{2}}\,,
\end{equation}
for $i=2,\ldots,n-2$ and for the highest level
\begin{equation}\label{BE3}
e^{i(\Phi_{a_n}-\Phi_{a_{n-1}})}\left(\frac{z_l^{I_{n-1}}+\sfrac{1}{2}}{z_l^{I_{n-1}}-\sfrac{1}{2}}\right)^L=\prod_k\frac{z_l^{I_{n-1}}-z_k^{I_{n-2}}-\sfrac{1}{2}}{z_l^{I_{a-{n-1}}}-z_k^{I_{n-2}}+\sfrac{1}{2}}\prod_{k\neq l}\frac{z_l^{I_{n-1}}-z_k^{I_{n-1}}+1}{z_l^{I_{n-1}}-z_k^{I_{n-1}}-1}\,.
\end{equation}
Equations
\eqref{BE1}-\eqref{BE3} are exactly the Bethe equations for the
compact $\mathfrak{gl}(n)$ symmetric spin chain. It is obvious that
there are $n!$ alternative forms of the above Bethe Ansatz equations,
corresponding to $n!$ permutations of the elements of the set $I$, which in turn are associated with the $n!$ different bottom-to-top paths on the Hasse diagram.

To conclude our new solution procedure for the $\alg{gl}(n)$-spin chain we give the
expression for the eigenvalues of \eqref{sln-ham}.
It only involves the roots $z^{I_{n-1}}$ on the last-level 
\begin{equation}\label{energy.formula}
E=2\sum_{k=1}^{m_{I_{n-1}}}
\frac{1}{\frac{1}{4}-\left(z_{k}^{I_{n-1}}\right)^2}\,,
\end{equation}
where $m_{I_{n-1}}$ is the number of roots of the eigenvalue
$Q_{I_{n-1}}(z)$, 
which according to \eqref{Q-eigen} is equal to 
\beq
m_{I_{n-1}}=m_{a_1}+m_{a_2}+\cdots+m_{a_{n-1}}=L-m_{a_n}.
\eeq
The derivation of the energy \eqref{energy.formula}
from the functional relations is given
in Appendix~\ref{app:energy}.

\section{Conclusions and Outlook}
\label{sec:conclusions}
In this paper we developed a novel, systematic procedure for constructing the
Baxter ${\bf Q}$-operators connected with the $\alg{gl}(n)$-spin chain
\eqref{sln-ham} with quasi-periodic boundary conditions. For illustration purposes we confined ourselves to the case where the quantum space is a multiple tensor product of the compact, fundamental representation of $\alg{su}(n)$.
The ${\bf Q}$-operators are constructed as transfer matrices associated with
infinite-dimensional representations of the Yangian $Y(\alg{gl}(n))$
built from Fock representations of the harmonic oscillator algebra. This involves rather simple, but hitherto unknown solutions \eqref{Lcanon} of the Yang-Baxter
equations with first order dependence on the spectral
parameter. These solutions provide fundamental building blocks for all
other required solutions via the standard fusion procedure. As a
result we derived the full set of functional relations, which enabled us to
obtain a new algebraic solution of the $\alg{su}(n)$-spin chain
independent of the Bethe Ansatz.

The construction we have presented in this paper generalizes in a beautiful fashion to compact spin chains with $\alg{su}(n|m)$ supersymmetry. In particular, the partonic Lax operators \eqref{boselax} naturally generalize to the supersymmetric case. Details on this will be reported elsewhere \cite{
Frassek:2010ga}.

Our methodology also generalizes to more general representations in
the quantum space: Recall that in this paper we always took it to be a
tensor product of $L$ fundamental $n$-dimensional
representations. Particularly relevant is the case of non-compact spin
chains. It will be interesting to spell out the exact relation between
our construction of $\Qop$-operators and a rather different approach
proposed in \cite{Derkachov:2010qe} and a large number of earlier work cited therein.


\subsection*{Acknowledgments}
We thank Volodya Kazakov, Zengo Tsuboi and Changrim Ahn for useful
discussions. One of us (VB) thanks Alexander Molev and Sergei Sergeev
for very illuminating discussions. 
When finishing this paper, we became aware of \cite{Kazakov:2010iu},
which also proposes an (apparently quite different) method for
constructing ${\bf Q}$-operators for compact $\alg{su}(n)$-spin
chains, based on the co-derivative formalism \cite{Kazakov:2007na}.
We would like to thank the {\it Institute for the Early Universe} at Ewha Womans University in Seoul, the {\it Center for Quantum Spacetime} at Sogang University in Seoul, and the APCTP in Pohang for hospitality during the completion phase of this work.
T.~{\L}ukowski is supported by a DFG grant in the framework of the SFB 647 {\it ``Raum - Zeit - Materie. Analytische und Geometrische Strukturen''} and by Polish science funds during 2009-2011 as a research project (NN202 105136).


\appendix

\section{Fusion of the Canonical \texorpdfstring{$\Lbf$}{}-operators}\label{app:fusion}
The calculations which one has to perform in order to prove formula \eqref{product2} are tedious but rather straightforward. It is enough to calculate the matrix product of the operator-valued matrices \eqref{Lop1} and \eqref{Lop2}, and subsequently use the similarity transforms \eqref{sim1} and \eqref{sim2}, keeping in mind the commutation relations between the various oscillators. For the convenience of the reader we present here the explicit action of these similarity transforms $\mathcal{S}_1$ and $\mathcal{S}_2$ on the oscillators.

\subsection*{Similarity transform \texorpdfstring{$\mathcal{S}_1$}{}}
\begin{eqnarray}
\mathcal{S}_1\oscbmo_{\dot a b}\mathcal{S}_1^{-1}=\oscbmo_{\dot a b}-\oscbpt_{\dot a b}\\
\mathcal{S}_1\oscbmt_{b\dot a}\mathcal{S}_1^{-1}=\oscbmt_{b\dot a}-\oscbpo_{b\dot a}
\end{eqnarray}
All other oscillators are unchanged.
\subsection*{Similarity transform \texorpdfstring{$\mathcal{S}_2$}{}}
\begin{eqnarray}
\mathcal{S}_2\oscbmo_{\dot a b}\mathcal{S}_2^{-1}&=&\oscbmo_{\dot a b}-\sum_{\ddot c\not \in I\cup J}\oscbpt_{\dot a \ddot c}\oscbmo_{\ddot c b}\\
\mathcal{S}_2\oscbmt_{\dot b \ddot a}\mathcal{S}_2^{-1}&=&\oscbmt_{\dot b\ddot a}-\sum_{ c\in I}\oscbpo_{c \dot b }\oscbmo_{\ddot a c}\\
\mathcal{S}_2\oscbpo_{b \ddot a}\mathcal{S}_2^{-1}&=&\oscbpo_{b\ddot a}+\sum_{\dot c \in  J}\oscbpo_{ b \dot c}\oscbpt_{\dot c \ddot a}
\end{eqnarray}
All other oscillators are unchanged.

\section{Proof of the Factorization Formula}\label{app:factorization}
In this appendix we prove that the ordered product of the $n$ different partonic Lax operators, each containing $n-1$ pairs of oscillators, indeed can be disentangled by a similarity transform $\mathcal{S}_{\mathcal{L}}$ into the product of the $\alg{sl}(n)$ invariant Lax operator $\mathcal{L}^+$ and a compensating matrix $\mathbb G_{\mathcal L}$.
Further details were discussed in section \ref{sec:intro}. Instead of using the prescription developed in section \ref{sec:fusion} to derive equation (\ref{sunfac}) inductively, we consider the following decomposition of the partonic Lax operators:
\begin{equation}
\label{zerlaxung}
\Lbf_a(z_a)= {\rm U}_a^{-1}\,\left( \begin{array}{ccccccc}
1 \quad & \, \, & \, \, & \, \,& \, \,&  \, \, & \, \,\\
\, \, &  \ddots\,\,&\, \, &  \, \,&  \, \,&\, \,& \, \, \\
\, \, &   \, \, & \; 1\, \, & \, \,&  \, \,&  \, \, & \, \,\\
\, \, &   \, \, &   \, \,& \specbaz_a -\frac{n-1}{2}&\, \,& \, \,& \, \, \\
\, \, &    \, \,&  \, \,&  \widetilde{\mathbf{D}}_{a+1,a} \, \,& 1 \quad&  \, \,& \, \,\\
\, \, &   \, \,&  \, \,& \vdots \, \,&   \, \,& \ddots \, \,& \, \,\\
\, \, &   \, \,&  \, \,& \, \widetilde{\mathbf{D}}_{\,n,a\phantom{+1}}\, \,&   \, \,& \, \,& \;1\, \,\\
\end{array} \right)\,{\rm U}_{a+1}\,,
\end{equation}
with $z_a=z+\lambda'_a$.

The matrices ${\rm U}_{a}$ do not depend on the spectral parameter. They are of the form
\begin{equation}
{\rm U}_a=\Bigg(\Identity+\sum_{1\leq b<c<
a}\oscb_{bc}e_{bc}\Bigg)^{-1}\Bigg(\Identity-\sum_{b=1}^{a-1}\sum_{c=a}^n\oscb_{cb}^{\dagger}e_{cb}\Bigg)\Bigg(\Identity-\sum_{a\leq
b<c\leq n}\oscb_{bc}^{\dagger}e_{bc}\Bigg).
\end{equation}
The entries in the column below the spectral parameter of the middle matrix on the right hand side of equation (\ref{zerlaxung}) take the values
\begin{equation}
\widetilde{\mathbf{D}}_{ab}=-{\oscb}_{ab}^{\dagger}-{\oscb}_{ab}+\sum_{c=1}^{b-1}{\oscb}_{ac}^{\dagger}{\oscb}_{cb}+\sum_{c=a+1}^n{\oscb}_{ac}^{\dagger}{\oscb}_{cb}\,;
\end{equation}
the remaining non-diagonal entries vanish.

From this decomposition it is obvious that the product of the $n$ partonic Lax operators can be written as
\begin{equation}
\label{fullzerlaxung}
\Lbf_1(z_1)\cdots\Lbf_n(z_n)= {\rm U}_1^{-1}\,\left( \begin{array}{cccc}
\specbaz_1 -\frac{n-1}{2}\,\,  &\, \, & \, \,& \, \,\\
\widetilde{\mathbf{D}}_{2,1}\, \, &  \ddots\,\,&\, \, &  \, \, \\
\vdots\, \, &   \ddots\, \, & \ddots\, \, &  \, \,\\
 \widetilde{\mathbf{D}}_{n,1} \, \, &   \cdots\, \, & \;  \widetilde{\mathbf{D}}_{n,n-1}  \, \,&  \specbaz_n -\frac{n-1}{2}\,\,  \\
\end{array} \right)\,{\rm U}_{n+1},
\end{equation}
where two matrices ${\rm U}_1$ and ${\rm U}_{n+1}$ only contain an upper triangular part:
\begin{equation}
\begin{split}
{\rm U}_1=\,\left( \begin{array}{cccc}
1\,\,  &\;-\oscb_{12}^\dagger\, \, & \;\cdots\, \,&-\oscb_{1n}^\dagger \, \,\\
\, \, &  \;\ddots\,\,&\;\ddots\, \, &\vdots  \, \, \\
\, \, &   \, \, &\; \ddots\, \, &\;-\oscb_{n-1\,n}^\dagger  \, \,\\
\, \, &  \, \, &   \, \,& 1\,\,  \\
\end{array} \right); \quad\quad
{\rm U}_{n+1}=\,\left( \begin{array}{cccc}
1\,\,  &\;\oscb_{12}\, \, & \;\cdots\, \,&\oscb_{1n} \, \,\\
\, \, &  \;\ddots\,\,&\;\ddots\, \, &\vdots  \, \, \\
\, \, &   \, \, &\; \ddots\, \, &\;\oscb_{n-1\,n}\, \,\\
\, \, &  \, \, &   \, \,& 1\,\,  \\
\end{array} \right)^{-1}.
\end{split}
\end{equation}
Up to now, the lower triangular matrix in equation (\ref{fullzerlaxung}) contains all $n-1$ oscillators. A short calculation shows that half of them can be absorbed by the similarity transform $\mathcal{S}_{\mathcal L}=S_n\cdots S_1$ with
\begin{equation}
\begin{split} 
S_a&=\exp\left[\sum_{b=1}^{a-1}\left(\oscb_{ba}^{\dagger}+\sum_{c=b+1}^{a-1}\oscb_{ca}^{\dagger}\oscb_{bc}\right)\oscb_{ab}^{\dagger}\right].
\end{split}
\end{equation}
This transformation coincides with the one that can be obtained from the method\footnote{
As we are interested in the factorization formula where the compensating matrix appears on the right hand side of the Lax operator it is convenient to start from the partonic Lax operators $\Lbf_n(z_n)$ and multiply the remaining ones in the desired ordering from the left.}
established in section \ref{sec:fusion}. Furthermore, it leaves ${\rm U}_1$ invariant and its action on ${\rm U}_{n+1}$ factorizes as 
\begin{equation}
\mathcal{S}_{\mathcal{L}}\,{\rm U}_{n+1}\,\mathcal{S}_{\mathcal{L}}^{-1}={\rm U}_1^{-1}{\rm U}_{n+1}\,.
\end{equation}

As a result one obtains that the ordered product of $n$ partonic Lax operators can be written as
\begin{equation}
\label{fullzerlaxungtrans}
\Lbf_1(z_1)\cdots\Lbf_n(z_n)= \mathcal{S}_{\mathcal{L}}\,{\rm U}_1^{-1}\,\left( \begin{array}{cccc}
\specbaz_1 -\frac{n-1}{2}\,\,  &\, \, & \, \,& \, \,\\
{\mathbf{D}}_{2,1}\, \, &  \ddots\,\,&\, \, &  \, \, \\
\vdots\, \, &   \ddots\, \, & \ddots\, \, &  \, \,\\
 {\mathbf{D}}_{n,1} \, \, &   \cdots\, \, & \;  {\mathbf{D}}_{n,n-1}  \, \,&  \specbaz_n -\frac{n-1}{2}\,\,  \\
\end{array} \right)\,{\rm U}_1{\rm U}_{n+1}\,\mathcal{S}_{\mathcal{L}}^{-1},
\end{equation}
where
\begin{equation}
{\mathbf{D}}_{ab}=-{\oscb}_{ab}+\sum_{c=a+1}^n{\oscb}_{ac}^{\dagger}{\oscb}_{cb}.
\end{equation}
We identify the $\alg{sl}(n)$ invariant Lax operator in its factorized form (see e.g. \cite{Derkachov:2006})
\begin{equation}
\mathcal{L}^+\equiv{\rm U}_1^{-1}\,\left( \begin{array}{cccc}
\specbaz_1 -\frac{n-1}{2}\,\,  &\, \, & \, \,& \, \,\\
{\mathbf{D}}_{2,1}\, \, &  \ddots\,\,&\, \, &  \, \, \\
\vdots\, \, &   \ddots\, \, & \ddots\, \, &  \, \,\\
 {\mathbf{D}}_{n,1} \, \, &   \cdots\, \, & \;  {\mathbf{D}}_{n,n-1}  \, \,&  \specbaz_n -\frac{n-1}{2}\,\,  \\
\end{array} \right)\,{\rm U}_1
\end{equation}
and
\begin{equation}
\mathbb{G}_{\mathcal{L}}\equiv{\rm U}_{n+1}
\end{equation}


\section{Hasse Diagrams}
\label{App:Hasse}
\begin{figure}[t!]
\begin{center}
\begin{pspicture}(10,5)
\rput(5,0){\rnode{A0}{$\Qf_{\varnothing}$}}
\rput(3,2){\rnode{A1}{$\Qf_{\{1\}}$}}
\rput(7,2){\rnode{A2}{$\Qf_{\{2\}}$}}
\rput(5,4){\rnode{A12}{$\Qf_{\{1,2\}}$}}
\ncline[ArrowInside=->,ArrowInsidePos=1.0,nodesep=0.1]{A0}{A1}
\ncline[ArrowInside=->,ArrowInsidePos=1.0,nodesep=0.1]{A0}{A2}
\ncline[ArrowInside=->,ArrowInsidePos=1.0,nodesep=0.1]{A1}{A12}
\ncline[ArrowInside=->,ArrowInsidePos=1.0,nodesep=0.1]{A2}{A12}
\end{pspicture}
\end{center}
\caption{Hasse diagram for  $\mathfrak{gl}(2)$.}
\label{Hasse2}
\end{figure}
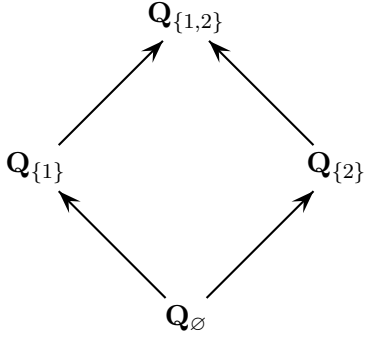

\begin{figure}[ht]
\begin{center}
\begin{pspicture}(10,9)
\rput(5,0){\rnode{A0}{$\Qf_{\varnothing}$}}
\rput(2,2){\rnode{A1}{$\Qf_{\{1\}}$}}
\rput(4,2){\rnode{A2}{$\Qf_{\{2\}}$}}
\rput(6,2){\rnode{A3}{$\Qf_{\{3\}}$}}
\rput(8,2){\rnode{A4}{$\Qf_{\{4\}}$}}
\rput(0,4){\rnode{A12}{$\Qf_{\{1,2\}}$}}
\rput(2,4){\rnode{A13}{$\Qf_{\{1,3\}}$}}
\rput(4,4){\rnode{A23}{$\Qf_{\{2,3\}}$}}
\rput(6,4){\rnode{A14}{$\Qf_{\{1,4\}}$}}
\rput(8,4){\rnode{A24}{$\Qf_{\{2,4\}}$}}
\rput(10,4){\rnode{A34}{$\Qf_{\{3,4\}}$}}
\rput(2,6){\rnode{A123}{$\Qf_{\{1,2,3\}}$}}
\rput(4,6){\rnode{A124}{$\Qf_{\{1,2,4\}}$}}
\rput(6,6){\rnode{A134}{$\Qf_{\{1,3,4\}}$}}
\rput(8,6){\rnode{A234}{$\Qf_{\{2,3,4\}}$}}
\rput(5,8){\rnode{A1234}{$\Qf_{\{ 1,2,3,4\}}$}}
\ncline[ArrowInside=->,ArrowInsidePos=1.0,nodesep=0.1]{A0}{A1}
\ncline[ArrowInside=->,ArrowInsidePos=1.0,nodesep=0.1]{A0}{A2}
\ncline[ArrowInside=->,ArrowInsidePos=1.0,nodesep=0.1]{A0}{A3}
\ncline[ArrowInside=->,ArrowInsidePos=1.0,nodesep=0.1]{A0}{A4}
\ncline[ArrowInside=->,ArrowInsidePos=1.0,nodesep=0.1]{A1}{A12}
\ncline[ArrowInside=->,ArrowInsidePos=1.0,nodesep=0.1]{A1}{A13}
\ncline[ArrowInside=->,ArrowInsidePos=1.0,nodesep=0.1]{A1}{A14}
\ncline[ArrowInside=->,ArrowInsidePos=1.0,nodesep=0.1]{A2}{A12}
\ncline[ArrowInside=->,ArrowInsidePos=1.0,nodesep=0.1]{A2}{A23}
\ncline[ArrowInside=->,ArrowInsidePos=1.0,nodesep=0.1]{A2}{A24}
\ncline[ArrowInside=->,ArrowInsidePos=1.0,nodesep=0.1]{A3}{A13}
\ncline[ArrowInside=->,ArrowInsidePos=1.0,nodesep=0.1]{A3}{A23}
\ncline[ArrowInside=->,ArrowInsidePos=1.0,nodesep=0.1]{A3}{A34}
\ncline[ArrowInside=->,ArrowInsidePos=1.0,nodesep=0.1]{A4}{A14}
\ncline[ArrowInside=->,ArrowInsidePos=1.0,nodesep=0.1]{A4}{A24}
\ncline[ArrowInside=->,ArrowInsidePos=1.0,nodesep=0.1]{A4}{A34}
\ncline[ArrowInside=->,ArrowInsidePos=1.0,nodesep=0.1]{A12}{A123}
\ncline[ArrowInside=->,ArrowInsidePos=1.0,nodesep=0.1]{A12}{A124}
\ncline[ArrowInside=->,ArrowInsidePos=1.0,nodesep=0.1]{A13}{A123}
\ncline[ArrowInside=->,ArrowInsidePos=1.0,nodesep=0.1]{A13}{A134}
\ncline[ArrowInside=->,ArrowInsidePos=1.0,nodesep=0.1]{A14}{A124}
\ncline[ArrowInside=->,ArrowInsidePos=1.0,nodesep=0.1]{A14}{A134}
\ncline[ArrowInside=->,ArrowInsidePos=1.0,nodesep=0.1]{A23}{A234}
\ncline[ArrowInside=->,ArrowInsidePos=1.0,nodesep=0.1]{A23}{A123}
\ncline[ArrowInside=->,ArrowInsidePos=1.0,nodesep=0.1]{A24}{A124}
\ncline[ArrowInside=->,ArrowInsidePos=1.0,nodesep=0.1]{A24}{A234}
\ncline[ArrowInside=->,ArrowInsidePos=1.0,nodesep=0.1]{A34}{A234}
\ncline[ArrowInside=->,ArrowInsidePos=1.0,nodesep=0.1]{A34}{A134}
\ncline[ArrowInside=->,ArrowInsidePos=1.0,nodesep=0.1]{A123}{A1234}
\ncline[ArrowInside=->,ArrowInsidePos=1.0,nodesep=0.1]{A124}{A1234}
\ncline[ArrowInside=->,ArrowInsidePos=1.0,nodesep=0.1]{A134}{A1234}
\ncline[ArrowInside=->,ArrowInsidePos=1.0,nodesep=0.1]{A234}{A1234}
\end{pspicture}
\end{center}
\caption{Hasse diagram for $\mathfrak{gl}(4)$.}
\label{Hasse4}
\end{figure}
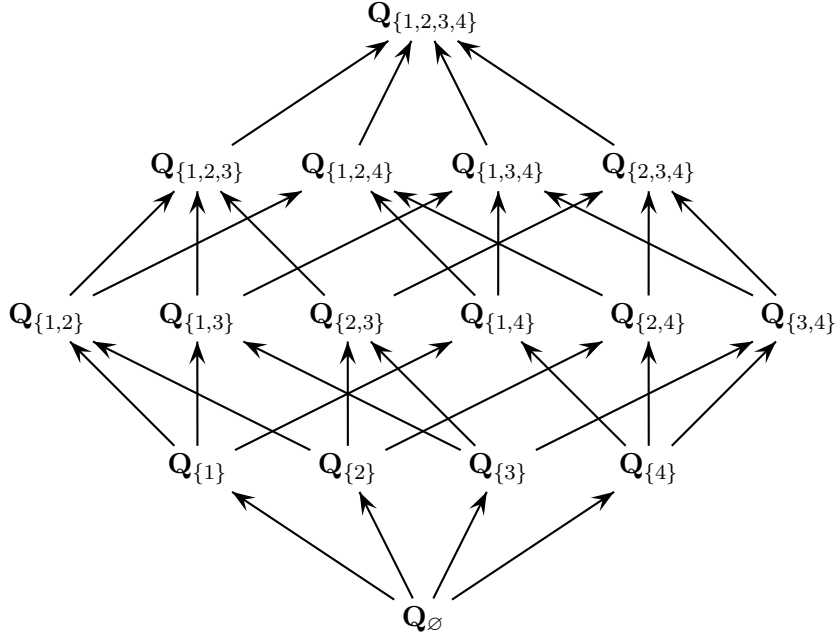

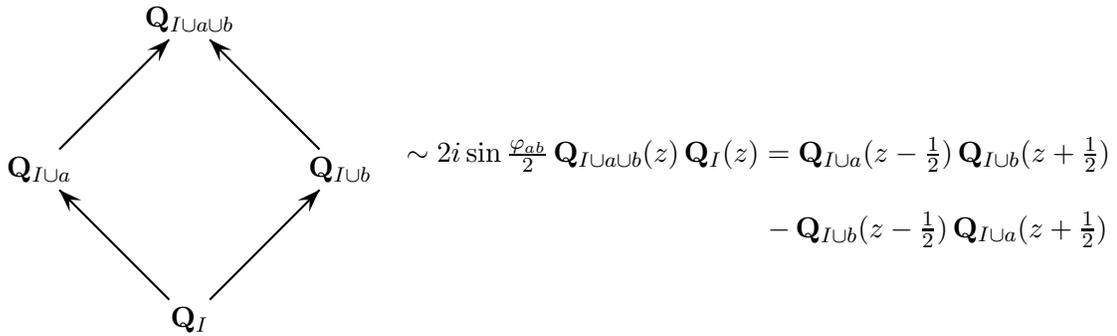
\begin{figure}[ht]
\begin{center}
\begin{pspicture}(15,5)
\rput(3,0){\rnode{A0}{$\Qf_I$}}
\rput(1,2){\rnode{A1}{$\Qf_{I\cup a}$}}
\rput(5,2){\rnode{A2}{$\Qf_{I\cup b}$}}
\rput(3,4){\rnode{A12}{$\Qf_{I\cup a\cup b}$}}
\ncline[ArrowInside=->,ArrowInsidePos=1.0,nodesep=0.1]{A0}{A1}
\ncline[ArrowInside=->,ArrowInsidePos=1.0,nodesep=0.1]{A0}{A2}
\ncline[ArrowInside=->,ArrowInsidePos=1.0,nodesep=0.1]{A1}{A12}
\ncline[ArrowInside=->,ArrowInsidePos=1.0,nodesep=0.1]{A2}{A12}
\rput(10.5,2.2){ $\sim 2i\sin\frac{\varphi_{a b}}{2}\,\Qop_{I\cup a \cup b}(\specbaz)\,\Qop_{I }(\specbaz)=
\Qop_{I \cup a}(\specbaz-\half)\,\Qop_{I \cup b}(\specbaz+\half)
$}
\rput(12.95,1.2){$ -\,
\Qop_{I \cup b}(\specbaz-\half)\,\Qop_{I \cup a}(\specbaz+\half)
$}
\end{pspicture}
\end{center}
\caption{Graphical depiction of the functional relations \eqref{FormJacobi}}
\label{Hasse0}
\end{figure}

The quadratic functional relations \eqref{FormJacobi} possess an
interesting graphical interpretation, cf.~Fig.~\ref{Hasse0}. The full
set of functional equations is then nicely depicted by so-called Hasse
diagrams, cf.~\cite{Tsuboi:2009ud} and references therein. These
diagrams are used to represent partially ordered sets. In our case we
take a power set of $\{1,\ldots,n\}$ with order given by the inclusion
relation, namely $A<B \Leftrightarrow A\subset B$. Then ${\Qop}_I$ will
inherit this ordering, giving us a partially ordered set containing
all $\Qop$-operators. By way of example, Hasse
diagrams for $n=2,3,4$ are presented in the figures
\ref{Hasse2}, \ref{Hasse3} and \ref{Hasse4}, respectively. To read off the
functional relations it is enough to take any 4-cycle in these
diagrams, using the equivalence depicted in figure \ref{Hasse0}.

Every path in the Hasse diagram which leads from ${\Qop}_{\varnothing}$ to ${\Qop}_{\{1,\ldots, n\}}$ defines a system of equivalent but distinct nested Bethe equations. To find each such system, it is enough to take all ${\Qop}$-operators on a given path and write one relation for any three subsequent functions on the path. Such relation can be written for every three subsequent ${\Qop}$'s because there always exists a unique 4-cycle containing them. Finally, it is interesting to point out that the Hasse diagram corresponding to the $\mathfrak{gl}(n)$ algebra forms an $n$-dimensional ordered hypercube.

\section{Energy Formula}
\label{app:energy}
Let us choose a path on the Hasse diagram given by the ordered set $\{a_1,\ldots,a_n\} = \{1,\ldots, n\}$. We define $I_{k}=I_{k-1}\cup \{ a_k\}$ with $I_0=\varnothing$.
The formula for the energy in terms of the Bethe roots depends only on the last-level ${\rm Q}$'s and is given by
\begin{equation}\label{energy.formula2}
E=2\sum_{k=1}^{m_{I_{n-1}}}\frac{1}{\frac{1}{4}-\left(z_{k}^{I_{n-1}}\right)^2}
\end{equation}
where $m_{I_{n-1}}$ is the number of roots of  the ${\rm Q}_{I_{n-1}}(z)$
function. To prove \eqref{energy.formula} it is enough to take the
functional relation
\begin{equation}\label{Plucker}
\frac{{{\rm X}}_{I_k}(z_0,z_2,\ldots,z_k)}{{{\rm X}}_{I_k}(z_1,z_2,\ldots,z_k)}=\frac{{\rm X}_{I_k}(z_0,z_1,\ldots,z_{k-1})}{{\rm X}_{I_k}(z_1,z_2,\ldots,z_k)}\frac{{\rm X}_{I_{k-1}}(z_2,\ldots,z_k)}{{\rm X}_{I_{k-1}}(z_1,z_2,\ldots,z_{k-1)}}+\frac{{\rm X}_{I_{k-1}}(z_0,z_2,\ldots,z_{k-1})}{{\rm X}_{_{k-1}}(z_1,z_2,\ldots,z_{k-1})}
\end{equation}
where ${\rm X}_{I_k}$ is an eigenvalue of $\Xbf_{I_k}$ defined in \eqref{Z-def}. It holds for $k=1,\ldots,n$ and originates from the fact that every ${\rm X}_{I_k}$ can be written as a determinant of partonic Q-functions as (see \eqref{Xasdet})
\begin{equation}
\Delta_{I_k}{{\rm X}}_{I_k}(z_1,\ldots,z_k)=\det_{a,b\in I_k} {{\rm Q}}_a(z_b)\,.
\end{equation}
In this case the relation \eqref{Plucker} is just a version of the Pl\"ucker relations and is valid for any numbers $z_a$. In \eqref{Plucker} we omitted an explicit dependence on the spectral parameter which can be recover for every $X_{I_k}$ just by taking the mean value of all $z$'s present there.

If we take $k=n$ and put $z_a=z-a+\frac{n+1}{2}$ then using this relation recursively we get
\begin{equation}
\frac{{\rm T}_{\Box}(z+\frac{1}{n})}{{\rm Q}_{I_n}(z)}=\sum_{k=0}^{n-1}\frac{{\rm Q}_{I_{n-k}}(z+1+\frac{k}{2})}{{\rm Q}_{I_{n-k}}(z+\frac{k}{2})}\frac{{\rm Q}_{I_{n-k-1}}(z-\frac{1}{2}+\frac{k}{2})}{{\rm Q}_{I_{n-k-1}}(z+\frac{1}{2}+\frac{k}{2})}\end{equation}
where ${\rm T}_{\Box}$ denotes an eigenvalue of the transfer matrix in the fundamental representation. In order to prove \eqref{energy.formula} it is enough to use the well known relation
\begin{equation}
E=2L-2\frac{d}{dz}\log {\rm T}_\Box(z+\frac{1}{n})\Big |_{z=0}
\end{equation}
and the fact that ${\rm Q}_{I_{n}}(z)=z^L$ where $L$ is the length of the spin chain.

\bibliographystyle{utphys}
\bibliography{bflms}
\end{document}